\documentclass[12pt]{article}

\usepackage{latexsym,amsmath,amssymb,theorem,epsfig}

\DeclareFontFamily{OT1}{rsfs10}{}
\DeclareFontShape{OT1}{rsfs10}{m}{n}{ <-> rsfs10 }{}
\DeclareMathAlphabet{\mathscript}{OT1}{rsfs10}{m}{n}

\numberwithin{equation}{section}

\newcommand{\ns}{\normalsize}

\newcommand{\RR}{{\mathbf{R}}}

\def\a{\alpha}
\def\b{\beta}
\def\g{\gamma}

\def\d{\delta}

\def\l{\lambda}
\def\m{\mu}

\def\r{\rho}
\def\s{\sigma}
\def\t{\tau}
\def\u{\upsilon}

\def\v{\varphi}

\def\D{\Delta}

\def\G{\Gamma}

\def\L{\Lambda}

\def\cj{{\cal J}}

\def\gsim{ \lower .75ex \hbox{$\sim$} \llap{\raise .27ex \hbox{$>$}} }
\def\lsim{ \lower .75ex \hbox{$\sim$} \llap{\raise .27ex \hbox{$<$}} }
\def\be{\begin{equation}}
\def\ee{\end{equation}}
\def\bea{\begin{eqnarray}}
\def\eea{\end{eqnarray}}

\def \td {\tilde}

\def \V  {{\rm V}}
\def \ha {{1 \ov 2}}

\def \sql {{\sqrt{\l}}\ }
\def \del{\partial}
\def \a {\alpha}
\def \aa {{\a'}}
\def\ov{\over}
\def \ci {\cite}

\def \foot {\footnote}
\def \bi{\bibitem}
\def\la{\label}\def\foot{\footnote}\newcommand{\rf}[1]{(\ref{#1})}

\def \no {\nonumber}

\def \adss {$AdS_5 \times S^5\ $}

\def \a {\alpha }

\def \V  {{\rm V}}

\theoremstyle{plain}

{\theorembodyfont{\rmfamily} }

\def \ed {\end{document}}

\begin{document}

%%%%%%%%%%%%%%%%%%%%%%%%%%%%%%%%%%%%%%%%%%%%%%%%%%%%%%%%%%%%%%%%%%%%%%

\begin{titlepage}

\vspace{-5cm}

%\hfill
%Imperial-TP-AT-2010-05 
%\vskip-1pt

\vspace{-5cm}

\title{
  \hfill{\small Imperial-TP-EIB-2012-01  }  \\[1em]
   {\LARGE 
   Correlation function of circular Wilson loop\\ with two 
   local operators 
    and conformal invariance
}
\\[1em] }
\author{
   E.I. Buchbinder\footnote{e.buchbinder@imperial.ac.uk} \ \  and 
     A.A. Tseytlin\footnote{Also at Lebedev  Institute, Moscow. 
   tseytlin@imperial.ac.uk }
     \\[0.5em]
   {\ns The Blackett Laboratory, Imperial College, London SW7 2AZ, U.K.}} %\\[-0.4cm]

\date{}

\maketitle

\begin{abstract}
We   consider the  correlation function of  a
circular Wilson loop with  two local  scalar   operators at generic 4-positions $a_1, a_2$ 
in planar ${\cal N}=4$ supersymmetric gauge theory.
We show that such  correlator is fixed by conformal invariance 
up to a function  $F({\rm u}, {\rm v}; \l)$ of 
 two scalar combinations ${\rm u}, {\rm v}$   of   $a_1, a_2$ coordinates
    invariant under the conformal transformations preserving the circle
  as well as  the 't Hooft coupling $\l$.
  We compute this  function  at  leading orders
at weak   and strong coupling  for  some simple choices 
of  local BPS operators.  We also check that correlators  of an infinite line Wilson loop
with local operators are the same as those for the circular loop. 

%The  leading order $\l$ term in the  correlator with  dimension 2   chiral primary 
%operators  involves  dilogarithm function. 

\end{abstract}

\thispagestyle{empty}

\end{titlepage}
%%%%%%%%%%%%%%%%%%%%%%%%%%%%%%%%%%%%%%%%%%%%%%%%%%%%%%%%%%%%%%%%%%%%%%
\def \u {{{\rm u}}}
\def \v   {{\rm {v}}}
\def \ed   {\end{document}}
\def \N {{\cal N}}
\def \RR {{\mathbb R}}
\def \tdr  {{\ell}} 
\def \r {r}  \def \L  {h}
\def \g {\gamma}
\def \ss {{\rm s}}
\def \ge {\varsigma}
 \def \OO {{\cal O}}

\def \Tr {{\rm Tr}}

 \def \adss {$AdS_5 \times S^5$\ }

\section{Introduction}

Supersymmetric Wilson loops \ci{mald} and their correlation functions 
  with local operators 
in planar   $\N=4$    SYM theory dual to $AdS_5 \times S^5$ string   theory is presently 
an active  subject of research. 
In this paper   we  will  focus on  correlators involving 
 the simplest  circular Wilson loop  
$W_C$  \ci{go, gog, corr, eric,dgr,sz}.
%\foot{We shall only  consider  the  planar approximation.}
 The  form of its    correlator 
 $\langle  W_C\  {\cal O} (a)\rangle $  with  one
   primary operator   ${\cal O}$ \ci{corr,z02}
 is completely fixed   by conformal invariance up to  a function
  of 't Hooft coupling $\l$   which may be computed 
exactly~\ci{sz, gpp} in the 
   case   when  the operator is  BPS.
 %   (chiral primary or its descendant).

 The  correlator  of $W_C$  with  {\it two}   chiral primary operators 
 can be again computed exactly \ci{gp} provided   their  
 locations  and structure 
 are special (so that at least  1/8   of supersymmetry is preserved
  \ci{gpp}).
Here   we  shall   consider a  ``non-supersymmetric''
  correlator  
 $\langle  W_C\  {\cal O}_1(a_1) {\cal O}_2(a_2)\rangle $ 
 with   {\it generic}  
  positions $a_1, a_2$ in ${\mathbb R}^4$   for  the simplest choices of 
  BPS operators ${\cal O}_i$.\foot{To  compare to \ci{gp}
  one would need  to consider the   special   operators  
  Tr$(a_k \Phi_k + i \Phi_4)^J$
   with coefficients depending
  on locations $a_k$  which  are  restricted  to the same   $S^2\subset 
  {\mathbb R}^4$   to which
  the circle  belongs. %This will not be attempted here.
 }
  As   we shall  find  below, the   conformal invariance     restricts the 
 dependence on  locations of the circular loop and the  two operators to 
 just two scalar functions $\u,\v$ of them, i.e. the  above 
  correlator is,  in  general,  proportional
   to a  function $F(u,\v;\l)$.
   We shall  compute this  function at leading orders at weak
  and strong coupling $\l$.
   
     %%%%%%%%%%%%%%%%%%%%%%%%%%%%%%%%%%%%%%
  
  The  circular   Wilson loop $W_C$   is known to be closely related to the Wilson loop $W_L$
  defined by an infinite straight line \ci{corr,dgr}.   Since 
  the  infinite   line   is related   to the   circle  
   by a special   conformal transformation % one may    expect
    the expectation 
    values of the two   would  be  the same   if not for   an  
anomaly~\ci{eric, sz, dgr}
    (related to  change  of boundary conditions).  Indeed,  $ \langle W_C  \rangle=1$  while 
    $\langle W_C  \rangle={ 2 \over \sql} I_1(\sql)$ is a nontrivial function of 
$\l$~\ci{eric, sz, dgr,gpp}.
    However, if one considers the {\it normalized} correlators  of $W_C$  with {\it local}
    operators 
    $  \langle W \, \OO_1(a_1) ...\OO_n (a_n)   \rangle \over  \langle W  \rangle $
    one may expect the anomaly to be  absent, i.e.  the result for  $W_C$ should be equivalent to the one 
    %v2
    for $W_L $.\foot{In the case of 1-point   correlator  $  \langle W \, \OO_1(a_1)    \rangle \over  \langle W  \rangle $
     this equivalence was suggested by N. Drukker  as mentioned  in \ci{fiol}.}
       This is  clear, in particular, at strong coupling  where  the expression for such correlator  (given by a product of the corresponding 
    vertex operators evaluated on the  minimal surface) 
    is finite  and thus  should not  be  affected by the anomaly. 
    %\foot{We thank N. Drukker and K. Zarembo  for a discussion  of this point.}  
    At weak coupling, one can  arrange the operators  to stay away from  the Wilson loop   location before   and 
     after the conformal transformation.
    Below  we  will explicitly will  check   the matching 
     $  \langle W_C\,  \OO_1(a_1) \OO_2(a_2)   \rangle \over  \langle W_C  \rangle $=
    $  \langle W_L\,  \OO_1(a_1) \OO_2(a_2)   \rangle \over  \langle W_L  \rangle $
    at leading order  in $\l$ for   simplest 1/2 BPS operators $\OO_i$.

 The    dependence  of   the  correlator  of the circle or line Wilson loop   with two local operators 
  on just two    invariants $(\u,\v)$ is reminiscent  of 
 the  familiar  structure of  the  correlator  of 4 scalar 
  conformal primary operators. Heuristically,  the fact   that an infinite 
    line   may be  specified   by  %a  point  and a  unit 4-vector of direction between  
    two points in ${\mathbb R}^4$    may be  suggesting  (by analogy with what was found in the 
    null polygon Wilson loop cases \ci{alk,abt,ada})
     a possible 
    relation between $\langle  W_C\  {\cal O}_1(a_1) {\cal O}_2(a_2)\rangle $
    and some special 4-point correlator.
    %We shall see indeed that the explicit  expressions  for $F(\u,\v)$ 
   % are reminiscent of the ones found for 4-point functions. 
  %  For example,  for the simplest choice 
 % of ${\cal O}_i$   as dimension 2 chiral primary operators 
 % the leading order-$\l$ contribution to  $F$ 
  % contains dilogarithm   function of $\u$.  
  % Logarithmic dependence of $F$ on $\u$ will be  found also at strong coupling.
 Another motivation for a  study of such correlators  is 
 that they  are  special  cases  of  correlators  involving  more general cusped  Wilson loops
 (see, e.g., \ci{gog,df,cor}).

 The  structure  of this  paper is as follows. 
 In section 2 we shall  consider   the conformal symmetry constraints on the 
 correlator of  a  circular Wilson loop with two  scalar conformal operators
 and explain   why it is determined by the function of two invariants 
 of the subset of 6 conformal transformations preserving the circular loop. 
 In section 3   we shall compute this function $F(\u,\v; \l)$  
 in the leading-order approximation at weak coupling for the case 
 when the two local  operators are  chiral primary  of dimension 2. 
 In section 4 we  shall  discuss  the 
 strong-coupling limit  of the  correlator $\langle  W_C\  {\cal O}_1(a_1) {\cal O}_2(a_2)\rangle $
   using the  semiclassical  string picture. 
 We shall   find that  for  two  ``light'' operators (whose dimension 
 does not scale with   $\sql$)  the correlator factorizes at strong coupling 
 with the function $F$  being constant. 
 In  the case   when one of the two operators carries 
 large ``semiclassical'' charge $J= \sql \cj $ the expression for $F$  
 will be  given by a non-trivial integral  which we shall  evaluate for small and large $\cj$. 
 
 In section 5 we shall discuss   the case of the Wilson loop  $W_L$  defined    by an  infinite line 
  and check the agreement of 
 its  correlator  with local operators    with the corresponding  correlators  for the circular 
 Wilson loop. 
 Some  technical  remarks   will  be made    in  Appendices A, B, C. 

%%%%%%%%%%%%%%%%%%%%%%%%%%%%%%%%%%%%%%%%%%%%%%%%%%%%%%%%%%%%%%%%%%%%%%%%%%%%%%%%%%%%%%%%%%%%%%%%%%%%%%%%%%%%%%%%%%%

\section{Conformal invariance constraints 
on 
correlator of\\ circular Wilson loop with two 
 scalar   operators   
}

%%%%%%%%%%%%%%%%%%%%%%%%%%%%%%%%%%%%%%%%%%%%%%%%%%%%%%%%%%%%%%%%%%%%%%%%%%%%%%%%%%%%%%%%%%%%%%%%%%%%%%%%%%%%%%%%%%%%%

In this section we  shall first  review  the  constraints 
on  some simplest  correlation functions in  ${\cal N}=4$   gauge  theory 
which follow from the conformal invariance  and then 
consider the case of 
$\langle  W_C\  {\cal O}_1(a_1) {\cal O}_2(a_2)\rangle $.

\subsection{Conformal invariance constraints  on some simple \\  correlation 
functions
 % in ${\cal N}=4$ gauge theory
}

Let us  start with 
correlation functions of scalar local 
operators ${\cal O}_i(a_i)$. 
As is  well-known, 
in conformal field theory  their 2- and 3-point functions  
are fixed by conformal invariance 
up to a constant (function of coupling)
while  a 4-point function  is  in general  proportional to a  function
of two cross-ratios (and coupling). 
%which, in general, depends on the 't Hooft 
%coupling $\lambda$. 
This can be seen, for example, as follows. 
Given a set  of $n$ points in ${\mathbb R}^4$
we can act on them with 15 generators of the conformal group. 
However, there can  be a subset of   generators 
which leaves this set  of points invariant. Let $\Gamma_0$ 
be the number of such generators. Then the number of conformally
 invariant combinations 
which one can construct out of  $n$ 4-coordinates  is
\be
d_n = 4n - (15-\G_0)\,. 
\label{1.1}
\ee
%
%where $4n$ is the total number of parameters in $n$ points. 
If $n=2$ we can place one point at the origin and
the other  at infinity. This configuration preserves dilatations 
and all the Lorentz transformations
which gives $\G_0= 7$. Then from~\eqref{1.1} we get $d_2=0$. This   means  
that one cannot construct any conformally 
invariant combinations and thus  the 2-point 
correlator is fixed up to a constant. As usual, the  latter can be fixed to 1 by a choice of
normalization,  i.e.  
%The operators are usually normalized 
%in such a way that this constant is unity % which gives the only possibility 
%
\be
\langle {\cal O}(a_1) {\cal O}^{\dagger}(a_2)\rangle =\frac{1}{|a_1-a_2|^{2 \Delta}}\,,
\label{1.2}
\ee
where $\Delta=\Delta(\lambda)$ is the dimension of the operator ${\cal O}$. 

%Now to the above system of two points we will add one more point 
%at some finite position. 
The $n=3$ case corresponds to adding 
an  extra point at some finite distance   from 0;  
that  breaks dilatations and  breaks Lorentz group  to  $SO(3)$. 
%of the Lorentz group will be  preserved.
 Hence for $n=3$ we get  $\Gamma_0= 3$ and $d_3=0$,  meaning 
that 3-point function is also fixed  
by conformal symmetry  up to a constant, i.e.  
is given by the well-known expression 
\be 
\langle {\cal O}_1(a_1){\cal O}_2(a_2){\cal O}_3(a_3)\rangle =
\frac{{\rm C}_{123}(\l) }{|a_1-a_2|^{\D_1+\D_2-\D_3}|a_1-a_3|^{\D_1+
\D_3-\D_2}|a_2-a_3|^{\D_2+\D_3-\D_1}}\,,
\label{1.3}
\ee
where $\D_i$ are  dimensions of ${\cal O}_i$.
%Thus, evaluation of 3-point functions is reduced to computations of
% the structure 
%constant ${\rm C}_{123}= {\rm C}_{123}(\lambda)$. 

Considering the $n=4$ case, i.e. adding one  more point 
at a finite distance from the origin 
% the system of 4 points (two of which are as before 
%at the origin and at infinity)
one finds that the remaining  symmetry is $SO(2)$, i.e.  $\G_0=1$ 
and thus $d_4=2$. This implies that 
the 4-point correlator is  fixed up to a function  $G$ 
of two conformally 
invariant variables %  (and $\l$)
%These variables are easy to construct in the form
%
\be
u= \frac{|a_1-a_2|^2 |a_3-a_4|^2}{|a_1-a_3|^2 |a_2-a_4|^2}\,, \qquad \ \ 
v= \frac{|a_1-a_4|^2 |a_2-a_3|^2}{|a_1-a_3|^2 |a_2-a_4|^2}\,.
\label{1.4}
\ee
The general expression for a  4-point function  may  then be written as\foot{There is,
 obviously, more 
than one way to  choose  
 the scaling prefactor  but  the ratio of any two   such 
prefactors is conformally 
invariant and  hence can be absorbed into  the function $G(u, v)$.}  
\be
\langle  {\cal O}_1(a_1){\cal O}_2(a_2){\cal O}_3(a_3){\cal O}_4(a_4)\rangle=
\frac{G(u, v; \l)}{|a_1-a_2|^{q_1}|a_1-a_4|^{q_2}|a_2-a_4|^{q_3}|a_3-a_4|^{q_4} 
}\,,
\label{1.5}
\ee
%
%where $G(u, v)$ is a function of the cross-ratios $u$ and $v$. 
$q_i$  are  fixed
by demanding that the correlator has dimension $\D_1+\D_2+\D_3+\D_4$ 
and that it  gets rescaled  by  $|a_1|^{2 \Delta_1} |a_2|^{2 \Delta_2}
|a_3|^{2 \Delta_3}  |a_4|^{2 \Delta_4} $ 
%
%\be
%\langle {\cal O}_1(a_1){\cal O}_2(a_2){\cal O}_3(a_3){\cal O}_4(a_4) \rangle 
%\to |a_1|^{2 \Delta_1} |a_2|^{2 \Delta_2}
%|a_3|^{2 \Delta_3}  |a_4|^{2 \Delta_4} 
%\langle {\cal O}_1(a_1){\cal O}_2(a_2){\cal O}_3(a_3){\cal O}_4(a_4) \rangle
%\label{1.3.1}
%\ee
%
under the inversions %. Since under inversions %$|a_i-a_j|$ goes to 
%
%\be
(when $|a_i-a_j| \to \frac{|a_i-a_j|}{|a_i| |a_j|} $)
%\label{1.3.2} 
%\ee
%
%we find that
%
\be
q_1= \D_1+\D_2+\D_3-\D_4\,, \ \ 
q_2= \D_1-\D_2-\D_3+ \D_4\,, \ \
q_3= - \D_1+\D_2 -\D_3 +\D_4\,, \ \ 
q_4 =2 \D_3 
\label{1.6}
\ee
%
%Thus, evaluation of 4-point functions is equivalent to 
%computing the function 
%$G(u, v)$ of two variables. Note that 

Let us now consider examples  of correlators 
of local  operators with  locally supersymmetric Wilson loop    
%Besides local operators there are other important observables in
%${\cal N}=4$ gauge thory. One of them is Wilson loop
\cite{mald}
\be 
W=\frac{1}{N} {\rm Tr} {\cal P}{\rm exp}\Big[\int d \tau (i A_{\mu} \dot{x}^{\mu} +
\Phi_I \theta_I |\dot{x}|)\Big]\,. 
\label{1.7}
\ee
Here 
$(A_{\mu}, \Phi_I)$ are bosonic  %gauge field and the scalars 
fields of ${\cal N}=4$ 
SYM  theory ($I=1,...,6$), $\theta_I \theta_I =1$ and 
$x^{\mu} =x^{\mu}(\tau)$ %is the equation of the
defines a  loop in ${\mathbb R}^4$.
 \iffalse 
  One can use $W$ to consider 
gauge invariant correlators of the form
%
\be 
\frac{\langle W {\ } {\cal O}_1(a_1){\cal O}_2(a_2)\ldots \rangle}
{\langle W \rangle}\,.
\label{1.8}
\ee
%
\fi
For example, in the case  of $W$   
%in~\cite{abt} the case of $W$ 
corresponding to the 
4-cusp null  polygon %was studied. 
it was shown in~\cite{abt} that the correlator 
%
%\be
$\frac{\langle W_4\, {\cal O} (a) \rangle}{\langle W_4 \rangle}
$
%\label{1.9}
%\ee
%
is fixed  by conformal invariance up to a function 
depending on a 
single invariant variable $\zeta$.\foot{This 
correlator   can thus be viewed 
as an ``intermediate'' case between the 3-point and 4-point functions of local
 operators.} 
% This makes this correlator one of the most 
%interesting 
%observables in ${\cal N}=4$ gauge theory and in the AdS/CFT correspondence.
% In fact it can be viewed
%as an intermediate case between 3-point and 4-point functions of local
% operators. 
%The fact that~\eqref{1.9} is fixed up to a function of a single variable 
%can be seen as follows.
Indeed,  let 
$x^{\mu(i)}$ ($i=1,2,3,4$) 
be  positions of the 4 cusps with $|x^{(i+1)}-x^{(i)}|=0$.
The total number of coordinates of 4+1 points is 20
but 4 null-line conditions reduce this number to 
16. % However, we have 
%4 null-line relations which gives only 16 independent variables.
Acting
 with 15 conformal generators 
leaves  only one conformally invariant combination\footnote{In this case
 $\G_0=0$. One can show 
that the 4-cusped null polygon is invariant under 3 conformal transformations.
 Addition of the operator(s)  
breaks all three of them.}
% \cite{abt}
%
\be
\zeta=\frac{|a-x^{(2)}||a-x^{(4)}| |x^{(1)}-x^{(3)}|}{|a-x^{(1)}||a-
x^{(3)}| |x^{(2)}-x^{(4)}|}\,,
\label{1.9.1}
\ee
 and the correlator has the following general form (see~\cite{abt} 
for details)
\be
\frac{\langle W_4 {\ } {\cal O} (a)\rangle}{\langle W_4 \rangle}=
\frac{\big(|x^{(1)}-x^{(3)}| |x^{(2)}-x^{(4)}|\big)^{\Delta/2}}{\prod_{i=1}^4
|a-x^{(i)}|^{\Delta/2}}\ {\rm F}(\zeta; \l)\,.
\label{1.10}
\ee
In~\cite{abt} the function ${\rm F}(\zeta;\l)$ was  found  to leading 
orders at weak and at strong coupling 
for ${\cal O}$ being the  dilaton and the chiral primary. 
Recently it was computed \cite{Alday:2012hy}
to the  next-to-leading order at weak coupling for the case of the dilaton
operator.  

In  determining the structure of \eqref{1.10}
%~\eqref{1.9.1}, \eqref{1.10}
 we assumed that the  conformal
 transformations 
act on the operator  as well as on the positions of the null cusps 
(in particular, $\zeta$ 
in~\eqref{1.9.1} is 
invariant under all  such conformal transformations).
 Alternatively, we can view the loop as a fixed object 
and consider  the correlation function %~\eqref{1.9} 
as a 
function of the position of the operator only.
 Then the positions
of the cusps are fixed  constants and we can  consider   simply  
$
\zeta^{\prime} =\frac{|a-x^{(2)}||a-x^{(4)}|}{|a-x^{(1)}||a-x^{(3)}|}
%\,. \label{1.10.1}
$
which %Unlike $\zeta$, the variable $\zeta^{\prime}$
 is invariant only under the  conformal
 transformations that 
preserve the null polygon.\foot{In~\eqref{1.10} we can also absorb the $a$-independent  numerator
 factor  %, which
 %is now viewed 
%as a constant independent of $a$, 
into the definition of ${\rm F}(\zeta^{\prime})$.}
Both approaches are  of course 
 equivalent. % and it is the matter of convenience which one to use. 

%%%%%%%%%%%%%%%%%%%%%%%%%%%%%%%%%%%%%%%%%%%%%%%%%%%%%%%%%%%%%%%%%%%%%%%%%%%%%%%%%%%%%%%%%%%%%%%%%%%%%%%%%%%%%%%%%%%%%%%%%%%%%%

\subsection{Correlator of circular Wilson loop and   one   operator}

%%%%%%%%%%%%%%%%%%%%%%%%%%%%%%%%%%%%%%%%%%%%%%%%%%%%%%%%%%%%%%%%%%%%%%%%%%%%%%%%%%%%%%%%%%%%%%%%%%%%%%%%%%%%%%%%%%%%%%%%%%%%

Another special  choice of $W$ is a circular Wilson loop $W_C$.
The correlator ${\langle W_C{\ } {\cal O}(a)\rangle}$
with one   local   operator 
%
%\be 
%\frac{\langle W_C{\ } {\cal O}(a)\rangle}{\langle W_C \rangle} 
%\label{1.10.2}
%\ee
%
also belongs to the class of 
simplest correlation functions:
 %In fact, at a given coupling 
it is fixed by conformal 
invariance up to a constant (function of $\l$)~\cite{corr,Gomis:2008qa,Alday:2011pf}.
This can  be seen  again by counting  the free parameters. 
It is convenient to view the circle as a fixed object. 
For concreteness,  we will assume that the circle is  in the $(x_1, x_2)$-plane in $ \RR^4 $
with the center at the origin 
\be 
x_1^2+ x_2^2 =R^2\,,\qquad  \qquad x_3=x_4=0\,.
\label{1.12}
\ee
As was shown in~\cite{Bianchi:2002gz} (see also Appendix A) that a circle in $ \RR^4 $ 
is invariant under 6 conformal transformations. The configuration of a circle and 
an operator preserves 6-4=2 of them. For example, if one places the operator 
at $a=\infty$  these 2 conformal transformations are a 
rotation in the $(x_1, x_2)$-plane and a 
rotation in the $(x_3, x_4)$-plane.
Then the number of combinations invariant under
the conformal transformations preserving the circle  is given by 
\be
d_{C, 1}= 4- (6- 2)=0\,. 
\label{1.11}
\ee
This formula is analogous to~\eqref{1.1} with  the dimension of the full conformal group 
 replaced with the  dimension of the subgroup preserving the circle. 
The fact that $d_{C, 1}=0$ means that we cannot construct any invariants and thus the correlation function of the 
circular Wilson loop and one local operator is fixed by the 
conformal invariance up to a constant (function of $\l$).
%rotation in the $(x_3, x_4)$-plane. Since $\G_0=2$ 
%we get from~\eqref{1.11} that $d_{C,1}=0$, i.e.  that  the above 
%correlator is fixed up to a constant (function of $\l$). 

The explicit  form of the correlator ${\langle W_C{\ } {\cal O}(a)\rangle}$
 can be found, e.g., by  using the fact that 
${\mathbb R}^4$ is conformal
 to $AdS_2 \times S^2$~\cite{Gomis:2008qa, Drukker:2005af}. 
 Let us write the metric
of ${\mathbb R}^4$ as 
\be
ds^2= dx_1^2 +dx_2^2 +dx_3^2 +dx_4^2 
=d\r^2+\r^2 d\psi^2 +d\L^2+\L^2 d\varphi^2\,, 
\label{1.13}
\ee
where $(\r, \psi)$ and  $(\L, \varphi)$ are the polar coordinates in the $(x_1, x_2)$  and  $(x_3, x_4)$
  planes. 
The circle~\eqref{1.12} 
is at $\r=R$, $\L=0$. 
Let us  transform to $AdS_2 \times S^2$, i.e. 
     change  from
$(\r, \psi, \L, \varphi)$ to $(\rho, \psi, \theta, \varphi)$ as  follows
\bea
\r=\tdr \sinh \rho\,, \quad    \quad \L= \tdr \sin \theta\,, 
\qquad 
\tdr\equiv \frac{R}{\cosh \rho - \cos \theta}=
 \frac{\sqrt{(\r^2+\L^2-R^2)^2+4 R^2 \L^2}}{2 R}\,.
\label{1.14}
\eea
In the new coordinates the metric becomes
\be
ds^2= \tdr^2\big( d \rho^2+ \sinh^2 \rho\ d \psi^2 + d \theta^2 
+\sin^2\theta\ d \varphi^2\big)= \tdr^2 ds^2_{AdS_2 \times S^2}\,.
\label{1.16}
\ee
Under this transformation the circular 
loop becomes the boundary of $AdS_2$ and, hence, is invariant 
under the isometries of $AdS_2 \times S^2$. 
Then if we compute the correlator  ${\langle W_C{\ } {\cal O}(a)\rangle}$ 
 in gauge theory defined on 
 $AdS_2 \times S^2$ it can be invariant under the  isometries only if it is a 
constant, i.e. 
\be
\frac{\langle W_C{\ } {\cal O}(a)\rangle}{\langle W_C 
\rangle}\Big|_{AdS_2 \times S^2}={\rm C}(\lambda)\,. 
\label{1.17}
\ee
To  transform this  back to ${\mathbb R}^4$
we note that under ~\eqref{1.14} %an operator of dimension $\D$ 
%transforms as 
we have  ${\cal O}(a) \to \tdr^{-\D}{\cal O}(a)$,  so that   
\be
\frac{\langle W_C{\ } {\cal O}(a)\rangle}
{\langle W_C \rangle}=\frac{{\rm C}(\lambda)}{[\tdr(a)]^{\D}}=
{\rm C}(\lambda)\Big[ \frac{4 R^2}{(\r^2+\L^2-R^2)^2+4 R^2 \L^2}
\Big]^{\D/2}\,, 
\label{1.18}
\ee
where  $\r^2 = a_1^2 + a_2^2$ and  $\L^2 = a_3^2 + a_4^2$
(here $a_\mu$   are the coordinates of the point $a$).  
Note that in the limit when the  position of the operator approaches 
a  point on the circle 
this correlator diverges
as $d^{-\Delta}$  where 
$d = \sqrt{ (\r - R)^2   + \L^2}$   is  the distance 
  between the point $a$  and a point  on the circle.
 % At the same time, it is regular when $a$  approaches the  center of the circle. 
 Also, \rf{1.18}  scales as  $ ( \r^2+ \L^2)^{-\Delta} = |a|^{-2\Delta}$
 in the limit when the  size of the circle goes to zero, 
 in agreement with  the OPE prediction  \ci{corr} (cf. \rf{1.2}).

For large $\l$ the coefficient ${\rm C}(\lambda)$  is, in general, 
  of order $\sqrt{\lambda}$ for  large $\l$. For example, for $\cal O$  being 
  the dilaton  operator  or chiral primary of fixed dimension $j$ 
 one gets \cite{corr} 
\be
{\rm C}_{dil}(\lambda)= \frac{\sqrt{6} \sqrt{\lambda}}{96 N}\,,
\ \ \ \ \ \ \ \ \ \ \ \ \ \ \
{\rm C}_{j}(\lambda)= \frac{\sqrt{j} \sqrt{\lambda}}{2^{j+1} N}\,.
\label{4.6}
\ee
For completeness, we present a derivation of these values   in Appendix C.

%
%Thus, evaluation of this correlator is reduced to finding the constant ${\rm C}(\lambda)$. 

%%%%%%%%%%%%%%%%%%%%%%%%%%%%%%%%%%%%%%%%%%%%%%%%%%%%%%%%%%%%%%%%%%%%%%%%%%%%%%%%%%%%%%%%%%%%%%%%%%%%%%%%%%%%%%%%%%%%%%%%%%%%

\subsection{Correlator of circular Wilson loop and two   operators}

%%%%%%%%%%%%%%%%%%%%%%%%%%%%%%%%%%%%%%%%%%%%%%%%%%%%%%%%%%%%%%%%%%%%%%%%%%%%%%%%%%%%%%%%%%%%%%%%%%%%%%%%%%%%%%%%%%%%%%%%%%%%%%%

Next, let  us  consider the  case   of our interest: the 
correlator of 
the circular Wilson loop~\eqref{1.12}
with two local operators 
\be
\frac{\langle W_C{\ } {\cal O}_1(a_1){\cal O}_2(a_2) \rangle}{\langle W_C \rangle} 
\label{2.1}
\ .  \ee
%
%and show that it is also fixed to a large extent by conformal 
%invariance. 
Let us again perform the counting of parameters. The   
 two operators give  4+4=8. In general, a configuration 
of a circle and two points is not invariant under any conformal transformations, i.e.  here  $\Gamma_0=0$.
Then the number of remaining  invariant parameters  is
\be 
d_{C, 2}=8-6=2   \la{33} \ , 
\ee
%
%The configuration of a circle and two points is specified by 
%9+4+4=17 parameters and is not,  
%This  configuration, 
%in general, 
% invariant under any 
%conformal transformations, i.e.  here  $\Gamma_0=0$. 
%Then the number of remaining invariant parameters  here is (cf. \rf{1.1},\rf{1.11})
%\be 
%d_{C, 2}=17-15=2   \la{33} \ , \ee
and, hence, 
the correlator~\eqref{2.1} is fixed by conformal symmetry up to a function of
two variables (functions of $a^\m_1,a^\m_2$  and location  of the circle) 
and the coupling $\l$. 
%which puts it in the same category as 4-point function of local operators. 
%In this   argument  we viewed~\eqref{2.1} 
%as a function not only of the positions of the operators but also
% of the circle
%and we let conformal transformations act on all the three  constituents of our system.
%Alternatively, we  may  view the loop as a fixed object and consider only the 
%conformal transformations that 
%preserve the loop. 
%In turns out that in the present case this approach is
% more convenient and from now on 
%we will consider the circle at fixed position 
%as in~\eqref{1.12}. Then the above argument 
% shows that the correlator~\eqref{2.1}
%is fixed by conformal invariance up to a function depending on two variables
These two variables, which we will denote as $\u$ and $ \v$, are invariant under 6 conformal 
transformations preserving the circle.
As we  shall now explain, % will see below that $\u$ and $u_2$
$\u$ and $ \v$ have a transparent geometric meaning. 

Let us perform the  change of coordinates~\eqref{1.14}, i.e.  consider the 
correlator~\eqref{2.1} in  a theory defined on 
$AdS_2 \times S^2$. % This correlator has to be invariant under 6 
%isometries of $AdS_2 \times S^2$:
Since  the circle is mapped to the boundary of $AdS_2$, it  is
 invariant under the 6 isometries of $AdS_2 \times S^2$, 
 and  the same should apply to the correlator, i.e. 
    the   isometries of $AdS_2 \times S^2$ are precisely 
the 6 conformal transformations 
which preserve the circle~\eqref{1.12} in ${\mathbb R}^4$. 
The natural 
two  functions of the coordinates $(a^\m_1,a^\mu_2)$  
invariant under the isometries 
of $AdS_2 \times S^2$ are the two geodesic distances between the two points: 
$\ss$ in $AdS_2$    and $\ge$ in  $S^2$.
Thus 
\be
\frac{\langle W_C{\ } {\cal O}_1(a_1){\cal O}_2(a_2) \rangle}{\langle 
W_C \rangle}\Big|_{AdS_2 \times S^2}= F(\ss, \ge; \l)\,. 
\label{2.3}
\ee
%
%where $\ell_1$ and $\ell_2$ are the geodesic distances in $S^2$ and  $AdS_2$ respectively. 
The two  invariants  $(\u,\v)$ of the conformal transformations 
from $SO(1,2) \times SO(3) \subset  SO(1,5)$  preserving the circle 
%which we shall denote as
%\foot{We use the same notation as for two
% cross-ratios  \rf{1.4} in the 4-point function 
%but there is of course no  obvious  connection.}  
%$\u$ and $\v$
 are then    some     functions  of $\ss$ and $\ge$, e.g., 
$\u= \cosh \ss $ and $\v= \cos \ge $.
 Given  the two points 
$(\rho_1, \psi_1, \theta_1, \varphi_1)$ and $(\rho_2, \psi_2, \theta_2,
 \varphi_2)$ in 
$AdS_2 \times S^2$  corresponding to $a_1$ and $a_2$ 
in $\RR^4$ via \rf{1.13},\rf{1.14}, i.e.
\be
(a^\mu_{i}) \to (\r_i, \psi_i, \L_i, \varphi_i) \to  (\rho_i, \psi_i, \theta_i, 
\varphi_i)\,, %\quad \quad   \quad i=1,2\,.
\label{2.4.1}
\ee
it is straightforward to construct the corresponding  geodesics 
distances (see Apendix B).\foot{For $S^2$  the geodesic distance 
 is   given   by   the ``law of cosines'' --  
 a theorem in  spherical trigonometry  relating the sides 
 and angles of spherical triangles. 
 %(see Appendix B).
 }
  Explicitly, one  finds 
%We will define $\u$ and $\v$ as follows
%
\bea
&&
\u= \cosh \ss = \cosh \rho_1 \cosh \rho_2 -\sinh \rho_1 \sinh \rho_2 
\cos (\psi_2-\psi_1)\,,\la{uuu} \\
&&\v= \cos \ge= \cos \theta_1 \cos \theta_2 +\sin \theta_1 \sin \theta_2 
\cos (\varphi_2-\varphi_1)\,, 
\label{2.5}
\eea
where from~\eqref{1.14} we have ($i=1,2$)
\bea
&&\sinh \rho_i= \frac{\r_i}{\tdr_i}=\frac{2 R \r_i}{\sqrt{(\r_i^2+\L_i^2
-R^2)^2+4R^2 \L_i^2}}\ , 
\no\\
&&\sin \theta_i= \frac{\L_i}{\tdr_i}=\frac{2 R \L_i}{\sqrt{(\r_i^2+\L_i^2
-R^2)^2+4R^2 \L_i^2}}\,.
\label{2.6}
\eea
Transforming  back to ${\mathbb R}^4$ we get (cf. \rf{1.17},\rf{1.18})
\be {\cal C}(W_C, a_1, a_2; \l)=
\frac{\langle W_C{\ } {\cal O}_1(a_1){\cal O}_2(a_2) \rangle}{\langle 
W_C \rangle}=
\frac{1}{\big[\tdr(a_1)\big]^{\D_1}\ \big[ \tdr(a_2)\big]^{\D_2}}\ F(\u, \v; \l)\,,
\label{2.4}
\ee
where $\D_i$ %and $\D_2$
 are the dimensions of  ${\cal 
O}_i$
and  we used that   $\tdr_i=\tdr(a_i)$
where $\tdr$ was defined in \rf{1.14}
and $\u,\v$   depend on $a_1,a_2$  
 according to \rf{1.13},\rf{1.14},\rf{2.4.1},\rf{2.5}.
 
Note that   as follows  from \rf{1.14}
\be
|a_1-a_2|^2= \frac{2 \big[\cosh(\rho_1- \rho_2)-\cos(\theta_1- \theta_2)\big]}
{(\cosh \rho_1 -\cos \theta_1)(\cosh \rho_1 -\cos \theta_2)}=
2 \tdr(a_1) \tdr (a_2)\ (\u- \v)\,,
\label{3.10}
\ee
According to 
 the  definitions in ~\eqref{2.5} we have  $\u \geq 1$
  and  $|\v| \leq 1$.
   The  values 
$\u = 1$, $\v =  1$  are achieved only when
 $\rho_1=\rho_2,\ \psi_1=\psi_2,\ \theta_1=\theta_2,
 \varphi_1=\varphi_2$, i.e.   when 
 the operators are at the  coincident points $a_1=a_2$.
 Hence  the OPE  limit 
$a_1 \to a_2$ is equivalent to $\u  \to 1,\   \v \to 1$.
% i.e.  $\u \to \v \to 1$. 

%AT
Another limiting  case  is when $\u = 1$  and  $\v =  -1$, 
corresponding, e.g.,  to $\rho_1=\rho_2=0, \ \psi_1=\psi_2$
and $\theta_1=\theta_2= {\pi \ov 2} , \ \varphi_2=\pi, \ \varphi_1=0$.\foot{We thank S. Giombi to drawing our attention to this
case.}
In this case $\r_1=\r_2=0, \  h_1=h_2=R$ (with $\ell_1=\ell_2= R$), 
 i.e. the two points 
are at the  poles of the  2-sphere for which the circle is the equator, 
i.e.   in  cartesian coordinates  we have 
\be a_1 = (0,0,R,0) \ , \ \ \ \ 
a_1 = (0,0,-R,0)  \ , \ \ \ \ \ \   \u=-\v=1 \ .  \la{31}\ee
This  case corresponds to  a supersymmetric configuration considered 
in \ci{gp}.

Let us note also that the limit   when the radius $R$ of the  circle   goes to 0
(or, equivalently,  the locations $a_i$ go to infinity) 
corresponds to $\rho_i \to0, \ \theta_i \to 0$, so 
that  again $\u \to 1, \ \v \to 1$. 
In this limit the Wilson loop can be represented as a sum of local operators 
\ci{corr}, i.e. one has 
$ W_C = \langle W_C \rangle [ 1 + \sum _k   c_k   R^{\Delta_k}   {\cal O}_k (0
)  + ....] $ so that %$ \langle W_C  {\cal O_1}  \rangle
the  first non-trivial term in the $R\to 0$ limit of the 
correlator \rf{3.1} 
 will be proportional to the  corresponding 3-point function. 

Below
we will  explicitly compute  the leading terms in 
 $F(\u,\v;\l)$ for some simple cases of ${\cal O}_i$
 at weak  and at strong coupling.

%%%%%%%%%%%%%%%%%%%%%%%%%%%%%%%%%%%%%%%%%%%%%%%%%%%%%%%%%%%%%%%%%%%%%%%%%%%%%%%%%%%%%%%%%%%%%%%%%%%%%%%%%%%%%%%%%%%%%%%%

%{\bf to add: Asymptotics of $F$ in the limit of 
%$R \to 0$;  points approaching circle -- 
%why $\u,\v$ are regular  --  what if points approach 
%circle separately -- this is another OPE}

%%%%%%%%%%%%%%%%%%%%%%%%%%%%%%%%%%%%%%%%%%%%%%%%%%%%%%%%%%%%%%%%%%

\section{ The  correlator $ \langle W_C{\ } {\cal O}_1(a_1){\cal O}_2(a_2) \rangle 
$  %of circular Wilson loop with two  local operators
  at weak coupling}

%%%%%%%%%%%%%%%%%%%%%%%%%%%%%%%%%%%%%%%%%%%%%%%%%%%%%%%%%%%%%%%%%%%%%%%%%%%%%%%%%%%%%%%%%%%%%%%%%%%%%%%%%%%%%%%%%%%%%%%%%%%

Let us now  consider the correlator
\be
{\cal C}(W_C, a_1, a_2; \l)
= \frac{\langle W_C{\ } {\cal O}_1 (a_1) {\cal O}_2 (a_2)\rangle}{\langle W_C\rangle}
\label{3.1}
\ee
at weak coupling $\l \ll 1$. 
 We will choose the operators to be the simplest chiral primaries
\be
{\cal O}_1(a_1)= c_2 {\rm Tr}[Z^2(a_1)]\,,\quad
 {\cal O}_2 (a_2)= c_2 {\rm Tr}[\bar Z^2(a_2)]\,,\qquad 
 Z=\Phi_1+ i \Phi_2\, , \quad c_2= \frac{4  \pi^2}{\sqrt 2 N}\ . 
\label{3.2}
\ee
%
%where the normalization coefficient is
%
%\be
%c_2= \frac{4 \pi^2}{\sqrt{2}N}\,.
%\label{3.2.1}
%\ee
%
For the unit-radius  circle   ($R=1$) %chosen as (for simplicity here we will 
%set the radius $R$ to be 1) 
\be
x^{\mu} (\tau)= (\cos \tau, \sin \tau, 0, 0)\,,\qquad \qquad |\dot{x}|=1 \ ,  
\label{3.4}
\ee
the  Wilson loop~\eqref{1.7} is given by 
\be 
W=\frac{1}{N} {\rm Tr} {\cal P}{\rm exp}\Big[g\int d \tau(i A_{\mu} \dot{x}^{\mu} +
\Phi_1)\Big]\,.
\label{3.3}
\ee
%
%and for simplicity we set the radius $R$ to unity. 
In \rf{3.3} we  assume that  the fields in the euclidean 
${\cal N}=4$  SYM Lagrangian
$ L =  { 1 \ov 2 g^2} ( \Tr F^2_{\mu \nu} + ...) $
 are rescaled  by 
 the gauge coupling constant $g$ so that  $g$ appears 
only in the vertices. The 't Hooft coupling is defined as
$\l =g^2 N$.
%in terms of $g$ as 
%
%\be 
%\lambda= \frac{g^2 N}{4 \pi^2}\,.
%\label{3.4.1}
%\ee
%
We will   use the following conventions for the $SU(N)$ generators  
\be
A_{\mu}=A_{\mu}^a T^a\,, \qquad \Phi_I=\Phi_I^a T^a\,,
 \qquad {\rm Tr}(T^a T^b)=\frac{1}{2}\d^{ab}\,, \qquad 
a, b =1, \dots,  N^2-1\,.
\label{3.5}
\ee
Then %In our convention
 the propagators have the form 
\be
\langle A_{\mu}^a(a_1)  A_{\nu}^b(a_2) \rangle =\frac{\d_{\mu \nu}
 \d^{ab}}{4 \pi^2 |a_1-a_2|^2}\,, \qquad
\langle Z^a(a_1) \bar Z^b(a_2) \rangle =\frac{\d^{ab}}{2 \pi^2 |a_1-a_2|^2}\,.
\label{3.6}
\ee
With the choice of $c_2$   in \rf{3.2} 
 the two-point function 
% $\langle {\cal O}_1(a_1){\cal O}_2(a_2)\rangle
%=\langle {\cal O}_1(a_1){\cal O}_1^{\dagger}(a_2)\rangle$ 
 is canonically normalized\foot{Below we will always consider only the
planar approximation, i.e. the leading order in  large $N$ expansion.} 
%\footnote{In this paper all correlators
%are studied in the planar (large $N$) limit.}
%
\be
\langle {\cal O}_1(a_1){\cal O}_2(a_2)\rangle=\frac{1}{|a_1-a_2|^4}\,.
\label{3.7}
\ee
We will choose the locations of the operators as  (cf. \rf{1.13}) 
\be
(a_{1}^\mu)= (\r_1, 0, \L_1, 0)\,, \qquad (a_{2}^\mu)= (\r_2, 0, \L_2, 0)\,,
\label{3.9}
\ee
i.e.   the angles in 
\rf{1.13} are  $\psi_i=0, \  \varphi_i=0$. 
In this case, the  variables  $\u$ and $\v$ in \rf{2.5}
(invariant under the conformal transformations preserving the circle) 
take    simple form
\be \u= \cosh (\rho_1-\rho_2) \ , \qquad \qquad 
\v= \cos (\theta_1-\theta_2)\ .
\label{3.9.1}
\ee
%
%Let us now compute~\eqref{3.1} to order $g^2$. 
The numerator of  \rf{3.1}   contains a trivial 
 disconnected contribution  
$\langle W_C{\ } {\cal O}_1 (a_1) {\cal O}_2 (a_2)\rangle
 \sim \langle W_C{\ } \rangle  
\langle {\cal O}_1 (a_1) {\cal O}_2 (a_2) \rangle$.
Since the 2-point function of  chiral primary   operators is not renormalized, 
this   disconnected part 
  coincides with the 
2-point function~\eqref{3.7} to all orders in $g$
%\foot{Since the 2-point function  is not renormalized,
%this is  the exact form of the disconnected part of the correlator.}
%
\be
{\cal C}_{disc} = {\cal C}_{0}= \frac{1}{|a_1-a_2|^4}\,.
\label{3.8}
\ee
%
%which is also the leading contribution at  small $g$. 
 Using \rf{3.10}  we  see that 
this expression can indeed be written in the form~\eqref{2.4}
\be
{\cal C}_{0}= \frac{F_0 (\u, \v)}{ [\tdr(a_1)]^2 [\tdr (a_2)]^2}\,,
 \ \ \ \ \ \ \  \ \ \  F_0 (\u, \v)= \frac{1}{4 (\u-\v)^2}\,.
\label{3.11}
\ee
%
%i.e. the leading order contribution to the function $F$ in \rf{2.4}   is
%
%\be
%F_0 (\u, \v)= \frac{1}{4 (\u-\v)^2}\,.
%\label{3.12}
%\ee
%
The first non-trivial (connected)  contribution to ${\cal C}$ in \rf{3.1} 
 starts at order $g^2 \sim  \lambda$
\be
{\cal C}_1=\frac{g^2 c_2^2}{4 N} \int_{0}^{2 \pi} d\tau_1 \int_{0}^{\tau_1} d\tau_2
\langle {\rm Tr} [Z(\t_1) \bar Z(\t_2)  + Z(\t_2) \bar Z(\t_1)]
{\rm Tr}[Z^2(a_1)]{\rm Tr}[\bar Z^2(a_2)] \rangle_c \,,
\label{3.15}
\ee
where $ \langle... \rangle_c$ stands for connected  part 
of the correlator (here  computed in free-theory approximation).
%Here it is assumed that the fields in the loop are contracted with the operators. 
%Performing the necessary contractions we end up with 
As a result, 
\bea
{\cal C}_1%^{(2)}(W_C, a_1, a_2)
=\frac{g^2 N c_2^2}{64 \pi^6 |a_1-a_2|^2}
 \int_{0}^{2 \pi} d\tau_1 \int_{0}^{\tau_1} d\tau_2
%\nonumber \\ &&
\Big[ \frac{1}{|x(\t_1)-a_1|^2 |x(\t_2)-a_2|^2 }+ (a_1 \leftrightarrow a_2)
 %\frac{1}{|x(\t_1)-a_2|^2 |x(\t_2)-a_1|^2 }
 \Big]\,.
\label{3.16}
\eea
Using that here $\frac{\r_i^2+\L_i^2+1}{2 \r_i}=\coth  \rho_i $
%
%\be
%\frac{\r_i^2+\L_i^2+1}{2 \r_i}=\coth  \rho_i\,,
%\label{3.17}
%\ee
%
%the first integral in~\eqref{3.16} can be written as
%
we get \be
 \int_{0}^{2 \pi}  \int_{0}^{\tau_1}\frac{  d\tau_1 d\tau_2 }{|x(\t_1)-a_1|^2 |x(\t_2)-a_2|^2 }=
\frac{1}{4 \r_1 \r_2} \int_0^{2 \pi} \frac{d \tau_1}{\coth \rho_1 -\cos \t_1} 
\int_0^{\tau_1} \frac{d \tau_2}{\coth \rho_2 -\cos \t_2}\,.
\label{3.18}
\ee
This  resulting expression  for this integral is found to be 
%can be written as
%
\be 
\frac{1}{4 \r_1 \r_2}\   2 \pi^2 \ \sinh \rho_1 \ \sinh \rho_2 \ .
% \big[{\cal F} (U) + {\cal F} (U^{-1})\big]\,, 
\label{3.21}
\ee
%%%%%%%%%%%%%%%%%OUT %%%%%%%%%%%%%%%%%%%%%%%%%%%%
\iffalse
where\foot{Here ${\rm Li}_2(z) =\sum^\infty_{k=1} {z^k \ov k^2}=  
 - \int^z_0 { dt \ov t} \log (1-t) $.}
 % $U$ is given by
\be
 {\cal F} (U)= \frac{3\pi^2}{2} +2 \log U \Big[  \log (1+U )  - \log (1-U )  \Big]
 - 4 {\rm Li}_2 (U) + {\rm Li}_2 (U^2)\ , 
\label{3.24} 
\ee
with%\foot{The choice of  $ U= \u+  \sqrt{ \u^2 -1}$  corresponds to 
%$U \to U^{-1} $ and  thus  leads to an equivalent expression for \rf{3.21}.} 
\bea
&& U= \sqrt{\frac{(\coth \rho_1 +1)
(\coth \rho_2-1)}{(\coth \rho_1  -1)( \coth \rho_2
 +1)}} =  e^{\rho_1- \rho_2}= {\rm exp} ( {{\rm arccosh}\ \u})  \ , 
\label{3.26}\\
&&\u=   { 1 \ov 2} (   U +   U^{-1} )  \ ,  \ \ \ \ \ \  {\rm i.e. }  \ \ \ \ \ 
U= \u  \pm  \sqrt{ \u^2 -1}    \  , \ \ \ \  U^{-1}= \u  \mp \sqrt{ \u^2 -1} \ . 
\eea
%Note that for $U>1$ the function ${\cal F}(U)$ is not obviously real but 
%  it is straightforward 
%to check that  its  imaginary part  cancels thanks to the identities
%
%\be
%{\rm Im}( \log (1-U)) =\pi\,, \quad {\rm Im}( {\rm Li}_2 (U))=-\pi \log U\,, \quad U>1\,.
%\label{3.22.3}
%\ee
%
In~\eqref{3.21} ${\cal F} (U)+ {\cal F} (U^{-1})$ looks like a complicated function
of $U$ (and, hence, of ${\rm u}$). However, one can show that this sum is a constant independent of $U$:
%
\be
{\cal F} (U)+ {\cal F} (U^{-1}) =2 \pi^2\,.
\label{3.26.1}
\ee
%%%%%%%%%%%%%%%%%%%%%%%%%%%%%%%%%%%%
\fi
%Since~\eqref{3.21} is invariant under $a_1 \leftrightarrow a_2$,
 %$U \leftrightarrow U^{-1}$
The second integral in~\eqref{3.16}  produces  the same contribution. 
%as~\eqref{3.21}. 
Using   that  according to 
\eqref{1.14} 
%the denominator in~\eqref{3.21} can be written as
%
\be
 \frac{4 \r_1 \r_2}{ \sinh \rho_1 \ \sinh \rho_2}  ={ 4 \tdr(a_1) \tdr(a_2)}  \
 ,  
\label{3.25}
\ee
and taking into account   the value of  $c_2$ in ~\eqref{3.2}
we  get %obtain   the  following result for the correlator in
for  \rf{3.16} 
\be
{\cal C}_1 
 =\frac{\l }{8 N^2  }\frac{1}{\tdr(a_1) \tdr(a_2)  |a_1-a_2|^2   }   =   \frac{\l }{16 N^2  }\frac{1}{ [\tdr(a_1)]^2 [\tdr (a_2)]^2}
\frac{1}{\u- \v}\  , 
\label{3.30}
\ee
where we also   used    the relation ~\eqref{3.10}.
Thus  the order $\l= g^2 N $ term  in  the function $F(\u,\v; \l)$ in \rf{2.4} 
is given by  
\be
F_1 { (\u,\v)}= %\frac{g^2 N c_2^2}{256 \pi^6}\frac{{\cal F}(U) +{\cal F}(U^{-1})}{u_2-\u}=
\frac{\lambda}{16  N^2}\frac{1}{\u-\v}\,.
\label{3.31}
\ee
%
% and~\eqref{3.4.1}.
%%%%%%%%%%%%%%%%%%%%%%%%%%%%%%%%%%%%%%%%%%%%%%%%%%%%%%%%%
Let us now  study some special limits of   this expression. 
One is  the OPE limit $a_2 \to a_1$. In general, in this limit
we have the following leading singularity % behavior
\be 
{\cal O}_1 (a_1)  {\cal O}_2 (a_2)\sim 
\frac{1}{|a_1-a_2|^{\delta}} %\sum \d_p
 \ k_3\ {\cal O}_3 (a_1) + \dots\,, \ \ \ \ \ \ \ \ \ 
  \delta = \D_1+\D_2 -\D_3\ , 
\label{3.32}
\ee
where ${\cal O}_3$ stands  for %(a linear combination of) 
an operator  (or a linear combination of operators) of lowest dimension
such that $k_3 \sim  \langle {\cal O}_1 (a_1) {\cal O}_2 (a_2) {\cal O}_3 (0) \rangle$
is non-zero.  
%From eq.~\eqref{1.3} it follows that 
%
%\be
%\delta = \D_1+\D_2 -\D_p \,, 
%\label{3.32.1}
%\ee
%
%where $\D_p$ is the dimension of  ${\cal O}_p$. 
Substituting \eqref{3.32} into \eqref{3.1} gives
\be
{\cal C}_1\Big|_{a_2 \to a_1}\  \to \ 
\frac{k_3}{|a_1-a_2|^{\delta }} 
\frac{\langle W_C{\ } {\cal O}_3 (a_1) \rangle}{\langle W_C\rangle}
= \frac{1}{[\tdr(a_1)]^{\D_3}}\frac{1}{|a_1-a_2|^{\delta}} \ k_3\ {\rm C}_3 (\lambda)\,,
\label{3.33}
\ee
where %$\D$ is the dimension of the operators ${\cal O}_p$ and
%in the last equality 
we  used  that the correlator of the circular Wilson loop 
with one local operator is fixed by conformal invariance as in~\eqref{1.18}.
\iffalse 
To study the $a_2 \to a_1$ limit of \rf{3.30} it is useful to rewrite it,  
using \rf{3.10},  as 
%
\be
{\cal C}_1=\frac{\lambda}{8  N^2}
 \frac{1}{\tdr(a_1)\tdr(a_2)} 
\frac{1}{|a_1-a_2|^2}\,.
\label{3.34}
\ee
%
%The limit $a_2 \to a_1$ corresponds to  $\u \to 1$ and 
%$U \to 1$  when ${\cal F}(U=1)= {\pi^2\ov 2}$. 
\fi
In the limit $a_2 \to a_1$~\eqref{3.30} becomes
\be
{\cal C}_1\Big|_{a_2  \to a_1}   \to \  
 \frac{1} {[\tdr(a_1)]^2}\ \frac{1}{|a_1-a_2|^2} \ \frac{ \lambda }{8 N^2}  \,.
\label{3.35}
\ee
Comparing \eqref{3.35} with  \eqref{3.33} we  conclude  that 
here $\delta =2$ and   $\Delta_3=2$. 
Thus the leading contribution in this limit should 
come from   operators of dimension $2$   which have 
non-zero 3-point function with ${\rm Tr}[\bar Z^2]$ and ${\rm Tr}[\bar Z^2]$.
One obvious  choice  is  a  non-BPS
 operator $O_3 = {\rm Tr} [ Z \bar Z] + ... $.
% (which to this order may be  interpreted as  the Konishi operator). 
%{\bf Why? why not say $\Tr Z \bar Z$ ?
Another option is  to consider ${\cal O}_3$ 
as a particular  case of    generic  dimension 2   chiral primary    operator 
\be
{\cal O}  \sim {\rm Tr}[(n_{ I} \Phi_I)^2]\,,\ \ \ \ \ \ \ \ \ 
 n \cdot n =0 \ , \ \ \ n \cdot \bar n =2\ , 
\label{3.36}
\ee
with  ${\cal O}_{1}\sim {\rm Tr}[Z^2]$ 
and ${\cal O}_{2}\sim {\rm Tr}[\bar Z^2]$  corresponding, respectively,  to 
 $ n_1=(1, i, 0, 0, 0, 0) $ and  $ n_2= \bar n_1= (1, -i, 0, 0, 0, 0) $. 
Since   $\langle  {\cal O}_{1}{\cal O}_{2} {\cal O}_{3} \rangle $ 
is proportional to $ (n_1 \cdot n_2) (n_1 \cdot n_3) (n_2 \cdot n_3)$
the necessary conditions  on $n_3$   are  $(n_3 \cdot n_1)\not=0, \  
(n_3 \cdot n_2)\not=0$. 
The  contribution   of the BPS operators to the OPE 
will dominate at higher orders as their dimension will not grow with $\l$.

Another   special limit is when  one of the  2 points, e.g.,  $a_1$,  approaches
a point on  the circle, i.e. for the choice 
of coordinates in  \rf{3.9} this corresponds to 
 $\r_1 \to R=1$, $\L_1  \to 0$.
 In this limit  $\ell(a_1) $    in \rf{1.14}   reduces to the distance 
 $  d(a_1) = \sqrt{  (r_1-1)^2  + \L_1^2} $   from $a_1$   to the point 
 $(1,0,0,0)$ on the circle 
 while $\u$ and $\v$ stay finite.  As  could be  expected, 
    the behaviour of the correlator
 \rf{3.1},\rf{3.30} in this limit ${\cal C}_1 \to [d(a_1)]^{-2}$ 
 is the  same as of the single-operator correlator 
 in \rf{1.18}.

 %AT
 
Yet another special   case  related to the supersymmetric configurations 
considered in \ci{gp} is when  the two points belong to the 2-sphere around  the
center of the circle, e.g., $a_1=(0,0,1,0), \ a_2=(0,0,-1,0),$ 
when $\u=1, \ \v=-1$  (see \rf{31}; here $R=1$, $\ell_i= {h^2_i + R^2 \ov 2 R} =1$). 
Then 
%$U=1$,\  ${\cal F}(U) = {\pi^2 \over 2}   $
%and thus  
from \rf{3.30} we get 
\be 
{\cal C}_1{}_{(S^2)}  =\frac{\l }{32 N^2 }    \  .  \la{3.222} \ee 
As one can check, this 
  agrees with the expression found  %from the  matrix model
  in eqs.~(4.42), (4.43) 
in \ci{gp}.\footnote{In  eqs.~(4.42), (4.43) of~\cite{gp}  one has  to set $J_1=J_2=2$, 
$A_1=A_2=\frac{1}{2}A=2 \pi$, $s_2=1$, take into account the normalization of the 
chiral primary operators and note that the Wilson loop in~\cite{gp}   was  defined without 
the $1/N$ prefactor in front  (i.e. there  to the  leading order  $\langle W \rangle =N (1 + ...) $.}

%{\bf  Extension to order $\l^2$ ? }

%%%%%%%%%%%%%%%%%%%%%%%%%%%%%%%%%%%%%%%%%%%%%%%%%%%%%%%%%%%%%%%%%%%%%%%%%%%%%%%%%%%%%%%%%%%%%%%%%%%%%%%%%%%%%%%%%%%%%%%%%%%%

\section{The correlator $ \langle W_C{\ } {\cal O}_1(a_1){\cal O}_2(a_2) \rangle $
at strong coupling}

%%%%%%%%%%%%%%%%%%%%%%%%%%%%%%%%%%%%%%%%%%%%%%%%%%%%%%%%%%%%%%%%%%%%%%%%%%%%%%%%%%%%%%%%%%%%%%%%%%%%%%%%%%%%%%%%%%%%%%
\def\V {{\rm V}}

Let us  now consider the correlator~\eqref{3.1} at strong coupling 
 using  the \adss string theory   representation 
 \be 
{\cal C}(W_C, a_1, a_2)=\frac{1}{\langle W_C \rangle} \int_C
 {\cal D} \{X\}\ e^{-I(\{X\})}\ \V_1(a_1)\ \V_2 (a_2) \, .
\label{4.1}
\ee
Here $I(\{X\})$ is the string action proporional to the
 tension $ T= \frac{\sqrt{\lambda}}{2 \pi}$
and in the planar approximation the path integral is performed 
over the Euclidean wordsheets 
with the topology of a disc and boundary conditions set by the loop  $C$.
Local gauge invariant operators
${\cal O}(a)$ are  represented by vertex operators 
``inserted'' at  the boundary 
point $a$ 
of $AdS_5$  %(at 4-point $a$)%~\cite{Polyakov:2001af, Tseytlin:2003ac}
\be 
\V(a)=\int d^2 \xi\  V(\{ X(\xi)\}, a)\,.
\label{4.0.1}
\ee
In the limit of large $\lambda$ the path integral~\eqref{4.1} is %, in general, 
 dominated 
by a classical solution with boundary 
conditions prescribed by the loop $C$ % the Wilson loop 
(and  possibly  also by   the
vertex  operators  if they   carry  large charges of the same order as string tension $\sim \sql$). 
Semiclassical correlators of circular  loop with one vertex operator were
discussed, e.g.,  in \ci{z02,many,gp,ena}. Correlators with two operators similar to \rf{4.1}   were   
studied  recently
in  \ci{Alday:2011pf,gp}. 

We shall start with  the  
 case when  the two operators are ``light'', 
i.e.  have charges  much smaller than $\sql$ so that they do not change  the
form of semiclassical  surface  that ends on the circular  loop  at the boundary. 
The leading term in the  correlator \rf{4.1} then  factorizes 
into a product of $ \langle W_C{\ } {\cal O}_1(a_1)\rangle $
and  $ \langle W_C{\ }  {\cal O}_2(a_2) \rangle $. 
We shall then  consider  a  less trivial  case when one of the two operators 
is ``heavy'', i.e. has dimension  $J \sim \sql$. 
In both cases the aim  will be to check the
 general structure of the correlator \rf{2.4}
and 
 to compute the leading strong coupling contribution to the function $F(\u,\v;\l)$.

%%%%%%%%%%%%%%%%%%%%%%%%%%%%%%%%%%%%%%%%%%%%%%%%%%%%%%%%%%%%%%%%%%%%%%%%%%%%%%%%%%%%%%%%%%%%%%%%%%%%%%%%%%%%%%%%%%%%%%%

\subsection{Case of two light operators}

%%%%%%%%%%%%%%%%%%%%%%%%%%%%%%%%%%%%%%%%%%%%%%%%%%%%%%%%%%%%%%%%%%%%%%%%%%%%%%%%%%%%%%%%%%%%%%%%%%%%%%%%%%%%%%%%%%%%%%%

In  this  case 
 dimensions $\D_1$ and $\D_2$ are fixed, i.e.  much less than $ \sqrt{\lambda}\gg 1$.
 Then  %the operators do not contribute to the stationary-point 
%equations of motion and
 the classical 
solution which dominates the path integral~\eqref{4.1}
is the surface in $AdS_5$ ~\cite{corr,gog,z02} ending on the circle~\eqref{1.12}\foot{Here we  change the notation compared to \rf{3.4}  and 
use $\sigma$  instead of $\tau$ to parametrize the  unit-radius  circle  ($R=1$).  $\tau$  is then the  second 
  world-sheet coordinate, i.e. $\xi=(\tau,\sigma)$.}
\bea
z=\tanh \tau\,,\quad x_1= \frac{\cos \s}{\cosh\tau}\,, \quad 
x_2= \frac{\sin \s}{\cosh\tau}\,, \quad x_3=x_4=0\,, 
\ \ \ \ \  
\tau \in [0, \infty)\,, \quad \s \in [0, 2\pi]
\label{4.0}
\eea
where %$z$ is the radial coordinate in
the  $AdS_5$  metric is $ds^2 =  z^{-2} ( dx^\m dx^\m + dz^2)$. 
Then eq.~\eqref{4.1} becomes
\bea
{\cal C}_{_{{\sql}\gg 1}}=
\int d \tau_1 d \s_1\ V\big(z(\t_1, \s_1), x^{\mu} (\t_1, \s_1)-a_{1 }^\mu\big) \  
\int d \tau_2 d \s_2\ V\big(z(\t_2, \s_2), x^{\mu} (\t_2, \s_2)-a_{2}^ \mu\big)\,, 
\label{4.2}
\eea
where $z(\t, \s), x^{\mu} (\t, \s)$ is the solution~\eqref{4.0}. 
Each integral in~\eqref{4.2} 
is the strong-coupling limit of the  correlation function of the 
circular loop with the corresponding local operator
\be 
\int d \tau_i d \s_i\ V\big(z(\t_i, \s_i), x^{\mu} (\t_i, \s_i)-a^\mu _{i}\big)=
\frac{\langle W_C {\ } {\cal O}(a_i)\rangle}{\langle W_C \rangle}\,, %\quad i=1,2\,.
\label{4.3}
\ee
i.e. if $\D_1, \D_2 \ll \sqrt{\lambda}$  
the correlator \rf{4.1}
factorizes in the  strong coupling  
 approximation\footnote{This strong-coupling 
factorization was also  observed   in~\cite{gp}.} 
\be
{\cal C}_{_{{\sql}\gg 1}}= \frac{\langle W_C {\ }{\cal O}(a_1)\rangle}{\langle W_C \rangle} 
\frac{\langle W_C {\ } {\cal O}(a_2)\rangle}{\langle W_C \rangle} \,.
\label{4.4}
\ee
%
%Note that this factorization is universal and takes place for all operators. 
%As was discussed in section 2,
Since  the correlation function of a circular Wilson 
loop with a local operator
is fixed by conformal invariance 
to have   the form~\eqref{1.18} we conclude, comparing to \rf{2.4}, 
that the function $F(\u, \v;\l)$    is  constant ($\u,\v$ independent) 
in this limit 
\be
\sqrt{\lambda}\gg 1: \ \ \ \ \ \   \ \ \ \  
F(\u, \v; \l) = {\rm C}_1(\lambda) \ {\rm C}_2(\lambda)\,.
\label{4.5}
\ee
Here $ {\rm C}_i(\lambda)$ is the corresponding coefficient in~\eqref{1.18}
(given explicitly in \rf{4.19} in the case when 
 the light operator is the dilaton  or the chiral primary of dimension $j$).

\def \R {{\rm R}}

%%%%%%%%%%%%%%%%%%%%%%%%%%%%%%%%%%%%%%%%%%%%%%%%%%%%%%%%%%%%%%%%%%%%%%%%%%%%%%%%%%%%%%%%%%%%%%%%%%%%%%%%%%%%%%%%%%%%%%%%%%%

\subsection{Case of one heavy and one light operator}

%%%%%%%%%%%%%%%%%%%%%%%%%%%%%%%%%%%%%%%%%%%%%%%%%%%%%%%%%%%%%%%%%%%%%%%%%%%%%%%%%%%%%%%%%%%%%%%%%%%%%%%%%%%%%%%%%%%%%%%%%%%

Let us  now  consider the case when one
of the two  operators (say ${\cal O}_1$) 
is chosen to be a ``heavy''
 chiral primary operator
with  dimension $\D_1 =J \sim \sqrt{\lambda}$  so that 
\be
%{\cal J}
\cj=\frac{J}{\sqrt{\lambda}}
\label{4.7.1}
\ee
 is fixed  in the  large $\l$ limit. 
 ${\cal O}_2$  will be choosen to be the dilaton operator
 whose dimension is
$\D_2=4 \ll \sqrt{\lambda}$. 
In the presence of  ${\cal O}_1$ (inserted at infinity) 
  the solution~\eqref{4.0}  is  modified to \cite{z02}
\bea
&&
z= e^{\cj \t} \Big[\sqrt{\cj^2+1}\ \tanh (\sqrt{\cj^2+1}\ \tau\ + q) -\cj\Big]\,, 
\nonumber \\
&&
x_1= \R(\t) \cos \s\,, \qquad x_2= \R(\t) \sin \s\,, \qquad x_3=x_4=0\,, 
\label{4.10} \\
&&
\R(\t)\equiv  \frac{\sqrt{\cj^2+1}\ e^{\cj \t}}{\cosh(\sqrt{\cj^2+1}\ \t +q)}\,, \qquad 
q\equiv  \log (\sqrt{\cj^2+1} +\cj)\,,
\ \ \ \ \ 
\phi= i \cj \t\,, \no 
%\label{4.10}
\eea
where $\phi$ is an angle of big circle in $S^5$  and as in  \rf{4.0} here 
$\tau \in [0, \infty)$, $\s \in [0, 2 \pi]$. 
The solution starts  at  $\tau=0$ as   the unit circle~\eqref{1.12}  (with $\R(0) =1$)  
and at  $\tau \to \infty$  approaches 
\be z \sim e^{\cj \t} \ , \ \ \ \ \ 
x^\mu \to 0 \ , \ \ \ \ \  \R(\tau)  \to 0\ , \ \ \ \  \  
\phi \sim i \cj \tau\,.
\label{4.11}
\ee
This  asymptotics  corresponds to the  chiral primary operator inserted at 
$z=\infty$, $x^{\mu}=0$, i.e. 
the solution \eqref{4.10}
``interpolates'' between the circle and the operator. 

The correlator \rf{4.1}  can be written as follows
\bea
{\cal C}_{J\sim \sqrt{\lambda}\gg 1} &=&
\frac{\langle W_C{\ } {\cal O}_J (a_1) {\cal O}_{dil} (a_2)\rangle }
{\langle W_C \rangle} 
\nonumber \\
&=&
\frac{\langle W_C{\ } {\cal O}_J (a_1) \rangle}{\langle W_C \rangle} \ 
\frac{\langle W_C{\ } {\cal O}_J (a_1) {\cal O}_{dil} (a_2)\rangle }
{\langle W_C   {\ }  {\cal O}_J (a_1)\rangle} \,.
\label{4.11.1}
\eea
The first factor is the correlation function of the circular loop with the
heavy operator  found in~\cite{z02} to be\footnote{Our expression for $\td{{\rm C}}_J(\lambda)$ 
differs from the one in~\cite{z02} 
by the factor $2^{-J}$ because our normalization of $\tdr$ in~\eqref{1.14}
involves an extra factor of $\frac{1}{2}$.} 
\bea
&&\frac{\langle W_C{\ } {\cal O}_J (a_1) \rangle}{\langle W_C \rangle}
 =\frac{\td{{\rm C}}_J}{[\tdr(a_1)]^J}\ , 
\label{4.11.2}\\ 
&&
\td{{\rm C}}_J =2^{-J} \exp \Big( {\sqrt{\lambda}\Big[1-\sqrt{\cj^2+1}-\cj
\log(\sqrt{\cj^2+1}-\cj)\Big]}\Big)
\,.
\label{4.11.3}
\eea
The second factor in~\eqref{4.11.1} is 
given by the  light vertex operator evaluated on 
the classical solution~\eqref{4.10}
\be 
\frac{\langle W_C{\ } {\cal O}_J (a_1) {\cal O}_{dil} (a_2)\rangle }
{\langle W_C   {\ }  {\cal O}_J (a_1)\rangle} =\int d \t d \s\
 V_{dil} \big(z(\t, \s), x^{\mu}(\t, \s)-a_{2}^ \mu, \phi(\t, \s)\big)\,.
\label{4.8}
\ee
%
%where $(z(\t, \s), x^{\mu}(\t, \s), \phi(\t, \s))$ is the solution~\eqref{4.10}, \eqref{4.10}.
Here  the dilaton vertex operator  is given by
\be
V_{dil}(a)= \hat{c}_{dil} \Big[\frac{z}{z^2+(x^{\mu}-a^{\mu})^2}\Big]^4 {\cal L} \ , 
%_{AdS_5 \times S^5}\,,
\ \ \ \ \ \ \ \ \ 
{\cal L}=\frac{(\partial_a z)^2 + (\partial_a x^{\mu})^2}{z^2} +(\partial_a \phi)^2 \ , 
\label{4.8.1}
\ee
where ${\cal L}$ is the \adss Lagrangian  in which  we ignored 
the extra  bosonic and fermionic coordinates that vanish on the classical solution.
 The normalization factor   $\hat{c}_{dil}$  is given by 
\be
\hat{c}_{dil}= \frac{\sqrt{6} \sqrt{\lambda}}{8 \pi N}\,.
\label{4.8.3}
\ee
To compute~\eqref{4.8} for general enough  values of $a_1,a_2$ 
(sufficient to restore the strong-coupling limit of the function $F(\u,\v;\l)$ in 
\rf{2.4}) 
we need the  classical solution corresponding 
to    the chiral primary operator  inserted at a finite point
on the $AdS_5$  boundary at $z=0$.
 It can   be found by  a conformal transformation  
 %AT
 applied to~\eqref{4.10}.\foot{Similar  conformal
 transformation was   considered in \ci{gp} and also in \ci{ena}.} 
Since the correlator under  consideration  is fixed to a large extent by the 
conformal symmetry  it is sufficient to  place
the operators at some special points $a_1,a_2$ as long as the variables $\u$ and
 $\v$ remain independent. 
We found the following choice to be convenient
% (here we list  both polar and cartesian coordinates, 
(see \rf{1.13}) 
\bea
a_{1}^{\mu}= (\r_1, \psi_1, \L_1, \varphi_1)= (0, 0, \L, 0)\,,  \ \ \ \ \ \ \ \ \ \ \ 
a_{2}^{\mu}= (\r_2, \psi_2, \L_2, \varphi_2)= (\r, 0, 0, 0)\,.
\label{4.12}
\eea
The chiral primary operator is  then located  above  the  center of the circle    while 
  the dilaton is inserted  in the plane of the circle (here $\r=1$  corresponds to a 
  point of the circle). 
In this case (see \rf{1.14},\rf{2.5},\rf{2.6})
\bea
{\tdr(a_1)}=\frac{1}{2}({\L^2+1})\,, \qquad {\tdr(a_2)}=\frac{1}{2}({\r^2-1})\,,
\  \ \ \ \ \ \ \ \ \ \u= \frac{\r^2+1}{\r^2-1} \ , \ \ \ \ \ 
\v= \frac{\L^2-1}{\L^2+1}\,.
\label{4.13}
\eea
%
%Hence, we have enough data to restore the function $F(\u, \v; \l)$ .
 Let us now perform a finite 
conformal transformation (an isometry of $AdS_5$) 
that  preserves  the circle   and  maps the point $(z=\infty, x^{\mu}=0)$ 
to the point 
$(z=0, x^\mu=a^\mu_{1})$. The transformation
consisting of a dilatation (with parameter $\gamma$), 
a special conformal transformation ($\beta_{\mu}$) and 
a translation ($\a_{\mu}$)   can be written as 
\be z' = \frac{\gamma  z }{1+2 \g \b \cdot x  +\gamma^2 \b^2 ( z^2+ x^2)} \ , \ \ \ \ \ \ \ 
x^{\prime}_{\mu}= \frac{\gamma  \big[ x_{\mu} + \g \beta_\mu (z^2 + x^2)\big] }{1+2\g  \b \cdot x
 +\gamma^2 \b^2 (z^2+ x^2)}+\a_{\mu}
  \,.
\label{4.14}
\ee
We will choose $\a^{\mu}= (0, 0, \a, 0)$, $\b^{\mu}= (0, 0, \b, 0)$.
Then the circle $x_1^2+x_2^2=1,\ x_3=x_4=0$  at $z=0$
%so that at $z=0$  
is transformed into 
%Then we get 
%
\bea
x^{\prime}_1=\frac{\g}{1+\g^2 \b^2}x_1\,, \qquad x^{\prime}_2=\frac{\g}{1+\g^2 \b^2}x_2\,,
\ \ \ \ \ \ \ \ 
x^{\prime}_3=\frac{\g^2 \b}{1+\g^2 \b^2} + \a \,, \qquad x'_4=0\,, 
\label{4.15}
\eea
so that to  preserve it  we have to require
\be
\frac{\g}{1+\g^2 \b^2}=1\,, \quad \frac{\g^2 \b}{1+\g^2 \b^2} + \a=0\ , \ \ \ 
{\rm i.e.} \ \ \ \   \a=-\sqrt{\g-1}\,, \ \ \ \  \b = { \sqrt{\g-1} \ov \g} \ . 
\label{4.17}
\ee
%
%The parameter $\gamma$ will be determined below. 
Note that this conformal transformation 
preserves the entire plane $x_3=x_4=0$. 
Acting   with \rf{4.14}  %this transformation
 on the solution~\eqref{4.10} 
we  obtain a new (conformally-equivalent)  solution\foot{The solution for $\phi$ is
 of course  unchanged and is still  given by~\eqref{4.10}.} 
\bea
&&
z'= \g w(\t)  z(\t)\,, \quad  x'_1=\g  w(\t) { \R(\t)  \cos\s}\,, \quad 
x'_2={\g   w(\t) \R(\t)  \sin\s}\,, \quad  x'_4=0\,, \\
&& x'_3=\sqrt{\g-1}   w(\t) { \big[z^2(\t) +\R^2(\t)-1\big]} \ ,\ \ \ \ 
 w(\t) \equiv \frac{1}{1+ (\g-1)\big[z^2(\t) +\R^2(\t)\big]}
\label{4.18}
\eea
where $z(\t)$ and $\R(\t)$ were given in~\eqref{4.10}. 
For  $\tau \to 0$ this solution still approaches 
the circle~\eqref{1.12}  while for  $\tau \to \infty$ we obtain
\be
z'=0\,, \qquad x'_1=x'_2=x'_4=0\,, \qquad x'_3=\frac{1}{\sqrt{\g-1}}\,.
\label{4.19}
\ee
To match the location $a_1$ of the  chiral promary operator  in \rf{4.12} we then need to fix 
 $\g$  as
\be
\g= \frac{\L^2+1}{\L^2}\,.
\label{4.20}
\ee
Let us now use  this  transformed 
 solution   to  compute the  contribution of the 
 light vertex operator in ~\eqref{4.8},\eqref{4.8.1}. 
Taking into account that the position of the dilaton operator is chosen as in 
 ~\eqref{4.12} and that the value of $ {\cal L}$ in \eqref{4.8.1} is 
\be
{\cal L} =\frac{2(1+\cj^2)}{\sinh^2 (\sqrt{\cj^2+1}\ \tau)}\ , 
\label{4.21} 
\ee
we can present~\eqref{4.8} in the form
\be
\frac{\langle W_C{\ } {\cal O}_J (a_1) {\cal O}_{dil} (a_2)\rangle }
{\langle W_C   {\ }  {\cal O}_J (a_1)\rangle} =
\frac{ \hat{c}_{dil}}{8(\cj^2+1)r^4} \int_0^{\infty}  d\tau \sinh^2 (\sqrt{\cj^2+1}\ \tau)
\int_0^{2 \pi} {d \s\over \big[y(\t) - \cos \s\big]^{4}}\,.
\label{4.23}
\ee
Here %$y(\tau)$ is 
\be
y(\tau)\equiv \frac{\g^2 (z^2+\R^2) + (\g-1) (z^2+\R^2-1)^2 +
r^2 \big[ 1+(\g-1)(z^2+\R^2)\big] ^2}{2 \g r \R \big[ 1+(\g-1)(z^2+\R^2)\big] }\,, 
\label{4.24}
\ee
with   $z=z(\tau)$ and  $\R=\R(\tau)$  given in~\eqref{4.10}.
Recall that  in view of \rf{4.13},\rf{4.20}    we have 
\be 
\r= \sqrt{ \u + 1 \ov \u-1 } \ , \ \ \ \ \  \ \ \ 
\L = \sqrt{ 1  + \v \ov 1 - \v  } \ , \ \ \   \ \ \ 
\g = { 2 \ov \v + 1 }   \ . \la{444}
\ee 
Doing the  integral over $\s$ 
%
%\be
%\int_0^{2 \pi} d \s (y- \cos \s)^{-4}= \frac{\pi y (3+2 y^2)}{(y^2-1)^{7/2}}\,,
%\label{4.25}
%\ee
%
 we end up with %the following non-trivial integral over $\tau$
\bea
&& \frac{\langle W_C{\ } {\cal O}_J (a_1) {\cal O}_{dil} (a_2)\rangle }
{\langle W_C   {\ }  {\cal O}_J (a_1)\rangle} = 
\frac{ \pi \hat{c}_{dil}}{8 (\cj^2+1) r^4}   \ I(\u, \v, \cj) \ , \la{4.260} \\
&&
I(\u, \v, \cj)= 
\int_{0}^{\infty} d \t \sinh^2 (\sqrt{\cj^2+1}\ \tau)
\ \frac{   [2 y^2(\t) + 3 ] \, y(\t)}{[y^2(\t) -1]^{7/2}}\,,
\label{4.26}
\eea
where we assume that $\r$ and $\g$ in $y$ 
are expressed  in terms of  $\u$ and $\v$  as in  \rf{444}.

Combining \rf{4.11.2} and  \rf{4.260} according to \rf{4.11.1}   and comparing to the 
general expression \rf{2.4} for the correlator  in question we  conclude that 
\be 
 {\l \gg 1}, \ \cj= { J \ov \sql} : \ \ \ \ \ \ \ \ \ \ \ \ 
  F(\u,\v; \l) 
 = \frac{ \pi\ {\td{{\rm C}}_J }\  \hat{c}_{dil} }{ 8(\cj^2+1) ( \u^2-1)^2 } 
   \ I(\u, \v, \cj)\ , 
\la{4.277}
\ee
where we used \rf{4.13} (i.e.  $ [{\tdr(a_2)}]^{-4} =16 ({\r^2-1})^{-4}$).
%and where $\r$ and $\g$ are related to $\u$ and $\v$  as in  \rf{444}.

%NEW

In the special case of ${\u}=1$, ${\v}=-1$ 
(see \rf{31}) corresponding 
here to $r \to \infty$, $\gamma \to \infty$ we get a finite expression for the function 
$F({\rm u}, {\rm v}; \lambda)$ in \rf{4.277}. Indeed,  in this limit
\be
y\    \to \   r  \frac{z^2+ {\rm R}^2}{{\rm 2R}}\,,
\label{Ad1}
\ee
 and then the $y$-dependent factor in the integrand of \rf{4.26}  becomes 
\be
\frac{(2 y^2 +3)y}{(y^2-1)^{7/2}}\  \  \to \  \ \frac{1}{y^4} \  \  \to \  \ 
%\frac{(u-1)^2}{(u+1)^2}
r^4 \Big(\frac{2 {\rm R}}{z^2+{\rm R}^2}\Big)^4\,.
\label{Ad2}
\ee
The  singular  factor  $r^4$ in  \rf{4.260}  then cancels out,  so that 
the correlator  becomes  a finite  constant (a function of ${\cal J}$ only).  
%Hence, we see from (4.32) that the limit ${\rm u} \to 1$ is finite. 

%AT
In general, the   integral  $I(\u, \v, \cj)$ in \rf{4.26}
appears to  be  too complicated to be  computable 
  analytically for arbitrary $\cj$ but it can be  easily 
  evaluated  in the  limiting cases of small and large $\cj$.

\subsubsection{Small  $\cj$  limit}

For $\cj=0$ the solution~\rf{4.10},\eqref{4.18}
becomes the original circle  solution~\eqref{4.0} and 
 ~\eqref{4.26} reduces to  the correlator  of 
the circular Wilson loop with the dilaton operator
%
%%%%%%%%%%%%%%%%%%%%%%%%%%%%%%%%%%%%%%%%%%%%%%%%%%%%%%%%%%%%%%%%%%%%%%%%%%%%%%%%%%%%%%%%%%%%%%%%%%%%%%%%%%%%%%%%%%%%%%%%%%
%\subsubsection{The limit $\cj=0$}
%%%%%%%%%%%%%%%%%%%%%%%%%%%%%%%%%%%%%%%%%%%%%%%%%%%%%%%%%%%%%%%%%%%%%%%%%%%%%%%%%%%%%%%%%%%%%%%%%%%%%%%%%%%%%%%%%%%%%%%%%%%
\be 
\cj \to 0: \ \ \ \ 
\frac{\langle W_C{\ } {\cal O}_J (a_1) {\cal O}_{dil} (a_2)\rangle }
{\langle W_C   {\ }  {\cal O}_J (a_1)\rangle} \ \ \to \ \ 
\frac{\langle W_C{\ } {\cal O}_{dil} (a_2)\rangle }
{\langle W_C   {\ }  \rangle}= \frac{{\rm C}_{dil}(\lambda)}{[\tdr(a_2)]^4}\,,
\label{4.27}
\ee
where ${\rm C}_{dil} $ was given in 
 ~\eqref{4.6}, i.e. 
 %and we used the fact that the dimension of the dilaton operator is 4. 
 in this limit the function $F(\u, \v;\l)$ is   constant  %given by 
\be{\l \gg 1, \   \cj \ll  1}: \ \ \ \ \ \ \ \ \ \ \ \ \ 
F(\u, \v; \l) %_{J\sim \sqrt{\lambda}\gg 1}= 
=\bar{{\rm C}}_J  {\rm C}_{dil}  \big[ 1 
 %(\lambda)  
% \frac{\sqrt{6} \sqrt{\lambda}}{96 N}
 +{\cal O}(\cj)\big]\,,
\label{4.28}
\ee
with 
 \be \bar{{\rm C}}_J=(\td{{\rm C}}_J )_{\cj \ll 1 } 
= 2^{-J} \exp \Big( \ha \sql \big[ \cj^2 +  O( \cj^4)\big]  \Big)   \ . \la{ccc} 
\ee
To find  the  linear in $\cj$ term in $F$  we 
expand the solution \rf{4.10} and thus $y$ in \rf{4.24}
 in powers of   $\cj$
\bea
&&z (\tau)=\tanh \t\big[1  +\cj  (\t- \tanh \t) + O(\cj^2)\big]\,, \ \ \  
\R (\tau)=\frac{1}{\cosh \t}\big[1 +{\cj}(\t -\tanh \t)+ O(\cj^2)\big] \,,
\label{n3} \no \\
&&
y(\tau)= \frac{1+r^2}{2r} \cosh\t + \cj 
\frac{(\g -2)(r^2-1)}{2 \g r} (\t \cosh \t -\sinh \t) + O(\cj^2)  \,. 
\label{n4}
\eea
Then the order $\cj$ term in \rf{4.260} becomes 
%Substituting it in~\eqref{n2} we obtain the following expression at order $\cj$
%
\bea
&&
\Big(\frac{\langle W {\ } {\cal O}_J (a_1) {\cal O}_{dil} (a_2) \rangle}
{\langle W {\ } {\cal O}_J (a_1)  \rangle} \Big)_{\cj}
=-16 \cj  \pi \hat{c}_{dil} \frac{\g-2}{\g}(r^2-1)\ {\cal I} (r)\,, 
\label{n5}
\\
%\nonumber\\
&&
{\cal I} (r)=
\int_{0}^{\infty} d \t \frac{\sinh^2 \t (\t \cosh \t- \sinh \t)}{\big[ (1+r^2)^2\cosh^2 \t -4r^2 
\big]^{9/2}}
\big[6r^4 +12 r^2 (1+r^2)^2 \cosh^2 \t + (1+r^2)^4 \cosh^4 \t\big]
\nonumber\\
&& \ \ \ \ \ \ \ = \frac{7+ 4r^2+2 r^4-12 r^6 -r^8+ 4 (1+r^2)^2 \log \frac{2r^2}{1+r^2}
}{12 (r^2-1)^6 (1+r^2)^3}\,.
\label{n6}
\eea
Expressing $\g$ and $r$ in terms of $\v$ and $\u$   according to \rf{444}, 
extracting the  factor $[\ell(a_2)]^{-4}$
and  also using  that   ${\rm C}_{dil}= {\pi \ov 12}\hat{c}_{dil}  $  (see \eqref{4.6})
%Note that using (4.31) the ratio $(2-\g)/\g$ is
% simply ${\rm v}$. Performing the integral we obtain
%
%\be
%{\cal I}=\frac{7+ 4r^2+2 r^4-12 r^6 -r^8+ 4 (1+r^2)^2 \log \left(\frac{2r^2}{1+r^2}\right)}{12 (r^2-1)^6 (1+r^2)^3}\,.
%\label{n6}
%\ee
%
%To find the function $F({\rm u}, {\rm v}; \lambda)$ we 
%need to pull out the factor $[\ell(a_2)]^{-4}$ and express
%$r$ in terms of $u$. We find 
%
we finally get   for the  order $\cj$ term  in  $F$ in \rf{4.28}
\be
F(\u, \v; \l) %_{J\sim \sqrt{\lambda}\gg 1}= 
=\bar{{\rm C}}_J  {\rm C}_{dil}  \Big[ 1 
 + \cj \frac{{\rm v}} {{\rm u}^3}
 \big( 1+2 {\rm u}^2-4 {\rm u}^3+ 4 {\rm u}^4 \log{ \u+ 1 \ov {\u}}
 \big)
 +{\cal O}(\cj^2)\Big]\,.
\label{4.288}
\ee

%%%%%%%%%%%%%%%%%%%%%%%%%%%%%%%%%%%%%%%%%%%%%%%%%%%%%%%%%%%%%%%%%%%%%%%%%%%%%%%%%%%%%%%%%%%%%%%%%%%%%%%%%%%%%%%%%%%%%%%%%%%%%%%%%%%%%%%%
\subsubsection{Large $\cj$ limit}

%{\bf To do:  expand in $\J$  and find leading correction -- this is of interest as interpolates to 
%case of 2 light operators}

%%%%%%%%%%%%%%%%%%%%%%%%%%%%%%%%%%%%%%%%%%%%%%%%%%%%%%%%%%%%%%%%%%%%%%%%%%%%%%%%%%%%%%%%%%%%%%%%%%%%%%%%%%%%%%%%%%%%%%%%%%%%%%%%%%%%%
In the limit of large $\cj$ one finds, to leading order, 
\be
z(\t)=\frac{1}{\cj} \sinh \cj \t\,, \qquad \R(\t)=1\,,
 \qquad y= \frac{1+r^2}{2r} +\frac{1+(\g-1)r^2}{2 \cj^2 \g r}\sinh^2 (\cj \t)\,.
\label{4.29}
\ee
Let us rescale $\cj \t \to \t$ and use $y$ as the new integration variable. 
Then up to terms subleading at large $\cj$ the integral~\eqref{4.26} can be written as
\be
I(\u,\v, \cj \gg 1) 
= 
\frac{\cj\g r }{1+(\g-1)r^2} \int_{\frac{1+r^2}{2r}}^{\infty} d y \ \frac{ 
(3+2 y^2)y}{(y^2-1)^{7/2}}\,.
\label{4.30}
\ee
The integral over $y$ gives
\be
\int_{\frac{1+r^2}{2r}}^{\infty} d y\ \frac{ (3+2 y^2)y}{(y^2-1)^{7/2}}=
\frac{16}{3} \frac{r^3 (1+4r^2 +r^4)}{(r^2-1)^5}\,.
\label{4.31}
\ee
As a result, from \rf{4.277} we get 
\be
{\l \gg 1, \   \cj \gg  1}: \ \ \ \ \ \ \ \ \ 
F(\u, \v;\l)= 
\frac{\hat {{\rm C}}_J  {\rm C}_{dil} }{2\cj}  \Big[ %\frac{\sqrt{6} \sqrt{\lambda}}{192  N}
\frac{3\u^2 - \v}{\u-\v} +{\cal O}({1 \ov\cj})\Big]\,,
\label{4.34}
\ee
where 
\be 
\hat {{\rm C}}_J = (\td{{\rm C}}_J )_{\cj \gg 1 } 
= 2^{-J} \exp \Big(  \sql \big[ \cj (\log (2\cj) -1 )  +
 1 +  O( \cj^{-1} )\big]  \Big)   \ . \la{c1c} 
\ee
Note that the leading singularity in the OPE limit $a_1 \to a_2$ is still $(\u-\v)^{-1}
 \sim |a_1-a_2|^{-2}$
just like at weak coupling (see \eqref{3.31}).
%\eqref{3.34}). 
%More precisely,~\eqref{4.33} in this limit is given by 
Explicitly, in this limit  %\eqref{4.33} is given by
\be
\Big({\cal C}_{\cj \gg 1}  \Big)_{a_1 \to a_2} \ \ 
%(W_C, a_1, a_2)_{J\sim \sqrt{\lambda}\gg 1} \ 
 \to  \   \   \frac{1}{[\tdr(a_1)]^{J+2}}\frac{1}{|a_1-a_2|^2}
\frac{2 \td{{\rm C}}_J {\rm C}_{dil} }{ \cj}\,, 
\label{4.35}
\ee
where we  used eq.~\eqref{3.10} and that in this limit $\u \to 1$, $\v \to 1$. 
Comparing with \eqref{3.33} we see that here  $\delta =2$ and that 
the leading contribution should come from  an  operator
of dimension $\D_3=J+2$.  This is   consistent with \rf{3.32},\eqref{3.33}
as we have  $\Delta_1=J,\  \Delta_2= 4$. 
%Also note that $h=2$ is consistent with eq.~\eqref{3.32.1} since in our present case
%$\D_1= J$, $\D_2=4$ and $\D$ has just been determined to be $J+2$. 

\def \aa {{a}}
%%%%%%%%%%%%%%%%%%%%%%%%%%%%%%%%%%%%%%%%%%%%%%%%%%%%%%%%%%%%%%%%%%%%%%%%%%%%%%%%%%%%%%%%%%%%%%%%%%%%%%%%%%%%%%%%%%%%%%%%%%%%%%

\section{Correlator of infinite line  Wilson loop \\
 with   local operators}

%%%%%%%%%%%%%%%%%%%%%%%%%%%%%%%%%%%%%%%%%%%%%%%%%%%%%%%%%%%%%%%%%%%%%%%%%%%%%%%%%%%%%%%%%%%%%%%%%%%%%

The locally-supersymmetric  Wilson loop \ci{mald,corr,gog}
 defined by an infinite straight  line (which we will denote as $W_L$)
  is a 1/2 BPS object with
   trivial expectation value, 
 $\langle W_L  \rangle=1$. 
If we choose the line    along the  $x_1$-direction,  i.e.  
\be
x_1=\tau\ , \ \ \ \ \ \ \ \ x_2=x_3=x_4=0\,,
\label{D1}
\ee
 the field  combination  in \rf{1.7}  becomes ``chiral'' ($iA_1 +  \Phi_1$).\foot{Note that   
the  expectation value of any function of $iA_1 +  \Phi_1$ 
over  the  gaussian measure  defined  by $L= (\del_\mu A_1)^2 +
 (\del_\mu \Phi_1)^2 + ...$
vanishes.}
 The  infinite   line \rf{D1} is related  \ci{corr,dgr}  to the   circle  
 \rf{1.12}  of radius $R$   with center at 0  by a particular 
 conformal transformation  (cf. \rf{4.14}) 
 %(and a translation to place the center at 
%a desired location)
%\footnote{For concreteness we performed the special conformal transformation 
%in the $x_2$-direction. To have the unit radius we have choose the parameter of the transformation
%$\b_2=\pm 1/2$.}
%
\be 
x_1^{\prime}=\frac{x_1}{1+\b^2 x_1^2}\,, \qquad x_2^{\prime}=\frac{\b x_1^2}{1+\b ^2 x_1^2}  -   R \,,
\ \ \ \ \ 
\beta\equiv \beta_2 = { 1 \ov 2 R} \ , \ \ 
\quad x_3^{\prime}=x_4^{\prime}=0\,,   
\la{d}
\ee 
where $x_1'^2 + x_2'^2 = R^2$.\foot{To get  the standard parametrization of the circle in \rf{3.4} 
we need also  to  change  $\tau\to \tau',\ \ \cos \tau'= {\tau \over 1 + {\tau^2 \over 4 R^2}}  $.} 
 The need to regularize (and the fact that the inversion changes   boundary conditions at infinity
 or changes topology of world surface on the string side) 
  lead  to an anomaly \ci{eric, dgr,sz}, 
 explaining why  the expectation value of the 
   circular Wilson loop  is no longer equal to 1:   its expression  is 
   given    in terms of the modified Bessel
   function of $\sql$,  
   $\langle W_C  \rangle={ 2 \over \sql} I_1(\sql)= 1 + {\l \ov 8} + {\l^2 \ov 192} + ...$. 
   
 As  was mentioned in the Introduction, one may  expect   that
despite $ \langle W_L  \rangle \not= \langle W_C  \rangle$ the transformation 
\rf{d}  may  still be    relating  the normalized 
correlators  of $W_L$ and $W_C$ with local operators, i.e.  the anomaly 
should be absent in the local correlators.

%It is then of interest to  compare the expectation values of the line
%Wilson loop with  local  operators to the  corresponding 
%correlators  with  the circular loop discussed  above.\foot{One may  expect, a priori,  that
%despite $ \langle W_L  \rangle \not= \langle W_C  \rangle$ the transformation 
%\rf{d}  may  still be  playing a role in  relating  normalized 
%correlators  of $W_L$ and $W_C$ with local operators. }

Let  us first discuss the  conformal symmetries preserved  by the
configuration involving a   straight line  \rf{D1}. 
As in the   circle case we may perform a conformal map from ${\mathbb R}^4$ 
 to $AdS_2 \times S^2$  with the line becoming the boundary 
 of $AdS_2$.
 %\foot{We may consider  instead   a  map to $S^4$ 
% with line becoming the  big circle of $S^4$.}
Here it is  natural to use the Poincare   coordinates for  $AdS_2$.
Explicitly,  going first to  spherical coordinates  in the $(x_2, x_3, x_4)$ subspace we get 
\bea
&&
ds^2= dx^2 + dz^2 + z^2 (d \theta^2 + \sin^2 \theta\ d \varphi^2)= 
 z^2 \big[ {dx^2 + dz^2 \ov z^2} +  ds^2_{S^2}\big] \,,
 \label{D2} \\
&&
x\equiv x_1\,, \qquad z=\sqrt{x_2^2+x_3^2+x_4^2}\,.
\label{D3}
\eea
An analysis similar to the one in Appendix A shows that the line~\eqref{D1} is preserved by  6 conformal 
transformations: dilatations, translations along the line, special conformal transformations along the line
and 3 rotations in the orthogonal space. These  may be interpreted  as the  isometries of $AdS_2 \times S^2$  
preserving the boundary (line $x$) of $AdS_2$. 

As  in the case of the circle, the correlation function of a line  with  one  local operator 
is fixed by conformal symmetry:  since the line is invariant under the  6  isometries it is impossible to construct 
an invariant  depending on  4-position of the operator, i.e. 
 by the same argument as in Section 2.2 we get (here
%\foot{Here 
$a^\mu$ are the cartesian coordinates of the point $a$  with the  direction of the line  being $x= a^1$) 
\be
\frac{\langle W_L {\ } {\cal O} (a)\rangle}{\langle W_L  \rangle}=
\frac{{\rm C}_L(\lambda)}{[\ell_L(a)]^{\Delta}}\,, \ \ \ \ \ \ \ \ \ \ \ 
\ell_L (a)\equiv  z= \sqrt{(\aa^2)^2+ (\aa^3)^2 +(\aa^4)^2}\,.
\label{D5}
\ee
Note that $\ell_L  (a)$ is just the  distance from the position of the operator to the line  \rf{D1}. 

%Beginning of <W_L O>/<W_L>
Let us compute    $\frac{\langle W_L { }{\cal O}(a) \rangle}{\langle W_L \rangle}$
to leading order in $\lambda$  for ${\cal O}$ being  a chiral primary operator 
and compare it with the corresponding  expression 
 for the circular Wilson loop. Using the definitions of the 
 Wilson loop  in~\eqref{3.3} and
the $\Delta=2$  operator  in~\eqref{3.2} we get for the order $\l$ term:
\be
\frac{\langle W_L {\ }{\cal O}(a) \rangle}{\langle W_L \rangle}= \frac{\lambda c_2}{32 \pi^4}
\int_{-\infty}^{\infty} d \tau_1 \int_{-\infty}^{\tau_1} d \tau_2 \frac{1}{|x(\tau_1)-a|^2\  |x(\tau_2)-a|^2} \,,
\label{n1}
\ee
where the  line  is  parametrized as $x(\tau)=(\tau, 0, 0, 0)$.
Performing the integrals  gives 
\be
\frac{\langle W_L {\ }{\cal O}(a) \rangle}{\langle W_L \rangle}= \frac{\lambda }{16 \sqrt{2} N}
\frac{1}{[\ell_L(a)]^2}  \,.
\label{n2}
\ee
This is the same result as found  in the case of the 
circle~\cite{sz}.\footnote{In ~\cite{sz}
the leading contribution at weak coupling is given in eq. (1.17).
The operator ${\cal O} $ which we used is 
$\frac{1}{\sqrt{2}}({\cal O}_{2}^1+i {\cal O}_{2}^2)$
in the notation of~\cite{sz}.} 
In general, one should have  (cf. \rf{1.18}, \rf{D5})
\be\la{ww}
[\ell_L (a)]^{-\D} \frac{\langle W_L {\ }{\cal O}(a) \rangle}{\langle W_L \rangle}= 
[\ell_C (a)]^{-\D}  \frac{\langle W_C {\ }{\cal O}(a) \rangle}{\langle W_C \rangle}, \  \ee
for all conformal operators and for all  values of $\l$. 

The exact expression for the correlator   \rf{n1}  is  found by replacing $\l$ in \rf{n2}  by  
$4 \sql {I_2(\sql)\ov I_1(\sql)} =  \l - {1\ov 24} \l ^2 + ... $  \ci{sz,gpp}.\foot{For 
dimension $k$ chiral primary 
one is to replace $I_2$ by $I_k$ \ci{sz}.} 
Since dimension 4  dilaton operator  is  in the same supermultiplet   with the $\D=2$ 
 chiral primary operator one may expect  that  its normalized 
 correlator with the circular   Wilson  loop 
 should   also be  proportional   to $ \sql {I_2(\sql)\ov I_1(\sql)}$. 
 This is indeed what one finds  if one  observes that\foot{In general, 
 $ x { d \ov d x }  I_k (x) = k + x { I_{k+1} (x) \ov I_k(x)}$.}
 \be 
 \sql  { d \ov d \sql} \log \langle W_C \rangle   =  \sql { d \ov d \sql} \log \Big[
   { 2 \over \sql} I_1(\sql)\Big] =   \sql {I_2(\sql)\ov I_1(\sql)} \ , \la{der} \ee
   and that   differentiating   $\langle W_C \rangle$  over 
   the coupling produces the insertion of the {\it integrated}  (over 4-space) 
   dilaton operator. The latter is  
   the  gauge theory action  if the correlator is understood in terms of the gauge theory path integral 
   or   the   string theory action  if it is defined in terms of the string  path integral.
   
   There is, however, a subtlety if one tries to use this  argument to 
    deduce  the  value  of the  coefficient  ${\rm C}_{dil}(\l)$  in  the local correlator \rf{1.18}:
  % $\frac{\langle W_C {\ }{\cal O}_{dil} (a) \rangle}{\langle W_C \rangle}$ 
  integrating \rf{1.18}  over the position $a$   
      one gets  (for $\D_{dil}=4$)  the     integral 
   $\int d^4 a  [\ell(a)]^{-4}\sim 
    R^4  \int^\infty_0  \int^\infty_0   { r dr \   h dh \ov [(r^2+h^2-R^2)^2 + 4h^2 R^2]^2}  $ 
    which is linearly    UV divergent ($\sim  R \int ^\infty_0{ dh \ov h^{2}}$)   at $h \to 0$. 
   As usual, this   UV  divergence is to be regularized away  to make the comparison  to \rf{der}
    possible.\foot{See  section  2.1 in \ci{rot}    for a related discussion of   integrated dilaton insertion into 
    correlation functions  where  one also  needs to introduce a UV cutoff  (see also \ci{cos}). 
     Note  that similar divergence is found at strong coupling if one simply  evaluates the  string 
    action on the corresponding minimal surface \ci{gog} (see also section  4 in \ci{rot}).}
    
    In the case of the line  where  $\langle W_L \rangle=1$ 
    the  analog of \rf{der}   vanishes    but this does   not imply that ${\rm C}_{dil}(\l)$ should vanish 
    (what would be in contradiction with \rf{ww}). Indeed, 
     the corresponding integral 
    $\int d^4 a  [\ell_L(a)]^{-4} = \int^\infty_{-\infty}  da_1  \int^\infty_{-\infty}   { d^3 {\vec a} \ov |\vec a|^4}$
    is now  not only   UV   but also   IR   divergent (along the infinite direction of the line). 
     Its    subtracted value   should be zero, thus 
    reconciling  the fact that  $ {d \ov d \sql}     \langle W_L \rangle=0$ with the expected relation 
     \rf{ww}.

%%%%%%&&&&&&&&&&&&&&&&&&&&&&&&&&&&&&&&&&&&&&&&&&&&&
Let us now  turn to the case of the  correlator of $W_L$    with two operators.
Like for the case of the circle, the  correlator of the line with two operators 
\be
{\cal C} (W_L, a_1, a_2)=\frac{\langle W_L {\ } {\cal O}_1 (a_1) {\cal O}_2 (a_2)\rangle}{\langle W_L {\rangle }}
\label{D7}
\ee
is also   fixed up to a  function of 8-6=2 variables $\u,\v$ 
related to   the geodesic distances in $AdS_2$ and $S^2$ (see  \rf{2.5})
which are invariant under the conformal transformations preserving the line.
Here 
the  variable ${\rm u}$ should be written 
 in terms of the Poincare coordinates. 
 Using the relation  between the global and the Poincare coordinates in $AdS_2$ (cf. \rf{1.16}) 
\be
\cosh \rho= \frac{1+x^2+z^2}{2 z}\,, \qquad \cos \psi = \frac{2 x}{\sqrt{(x^2+z^2-1)^2+4 x^2}}\,,
\label{D8}
\ee
we find that   for the two points in $AdS_2$ 
with coordinates $(x_1, z_1)$ and $(x_2, z_2)$  corresponding to 
$(\rho_1, \psi_1)$ and $(\rho_2, \psi_2)$ \foot{Note that 
 since this is 
 just a  coordinate transformation  in euclidean $AdS_2$  that  should not change geodesic distances
 the variables $\u$ and $\v$ are  actually the same in both cases.}
\be
{\rm u}=1 +  \frac{(x_1- x_2)^2+ (z_1-z_2)^2}{2 z_1 z_2}\,.
\label{D9}
\ee
Hence 
\be 
{\cal C} (W_L, a_1, a_2)=\frac{1}{[\ell_L(a_1)]^{\D_1}[\ell_L(a_2)]^{\D_2}}F_L({\rm u}, {\rm v}; \lambda)\,.
\label{D10}
\ee
%
%where $\D_1$ and $\D_2$ are the dimensions of the operators ${\cal O}_1$ and ${\cal O}_2$. 
Here $x_{1,2}$  are  first components   of $a_{1,2}$, i.e.
 $x_i = \aa^1_{i}$,  while  $\ell_L(a_i)= z_i= \sqrt{ (\aa^2_{i})^2 + 
 (\aa^3_{i})^2 +  (\aa^4_{i})^2 }$ and 
 the distance between the points $a_1$ and $a_2$ is again given by
 \rf{3.10}: 
\be
|a_1-a_2|^2= (x_1- x_2)^2+ z_1^2 + z_2^2- 2 z_1 z_2\ {\rm v}\ =\ 2 \ell_L(a_1) \ell_L(a_2)\ ({\rm u} -{\rm v})\ .  
\label{D11}
\ee
%
%where we have used~\eqref{D6} and \eqref{D9}. Note that it has the form as in eq. (2.25).
As an example,  let us compute the correlator~\eqref{D7} to leading order at weak coupling 
for the case when the operators are the chiral primaries in \rf{3.2}. 
The leading connected contribution is still given by \rf{3.16}  but with different integration limits
\be
{\cal C}_1= \frac{g^2 c_2^2}{64 \pi^6 |a_1-a_2|^2}\int_{-\infty}^{\infty} d \t_1
\int_{-\infty}^{\tau_1} d \t_2 \left[ \frac{1}{|x(\tau_1)-a_1|^2 |x(\tau_2)-a_2|^2}+
(a_1 \leftrightarrow a_2)\right]\,, 
\label{D12}
\ee
where in the present case of the line  % is now parametrized as 
$x(\tau)=(\tau, 0, 0, 0)$.
To find $F_L$ in \rf{D10}  it is  sufficient to make  
a special  choice of  coordinates of 
the points $a_1$ and $a_2$ (here  we list  the values of the $AdS_2 \times S^2$ coordinates, i.e. $a_i= (x_i,z_i, \theta_i,\varphi_i)$)
\be
a_1=(0, z_1, \theta_1, 0)\,, \qquad a_2=(0, z_2, \theta_2, 0)\,.
\label{D14}
\ee
%
%From eqs.~\eqref{D7.1}, \eqref{D9}, \eqref{D10} it is easy to see that this
%choice allows one to fully reconstruct the answer for the correlator. 
 In this case it is 
straightforward to evaluate the integrals in \rf{D12}  to obtain 
(see \rf{3.2},\rf{D11})
\be
{\cal C}_1= \frac{\lambda c_2^2}{64 \pi^6 |a_1-a_2|^2} \frac{\pi^2}{\ell_L(a_1)\ell_L(a_2)}=
\frac{\lambda}{16 N^2} \frac{1}{[\ell_L(a_1)]^2 [\ell_L(a_2)]^2} \frac{1}{{\rm u}-{\rm v}}\,.
\label{D15}
\ee
Thus  the leading-order term in $F_L$ in  \rf{D10} is 
\be
F_{1L}({\rm u}, {\rm v})=\frac{\lambda}{16 N^2}\frac{1}{{\rm u}-{\rm v}}\,,
\label{D16}
\ee
%
%We see that the function $F_{1L}({\rm u}, {\rm v})$ 
which is  the same as  $F_1$  in~\eqref{3.31} found 
for the circular Wilson loop.\foot{This agreement is not too surprising. As was  argued  in~\cite{dgr}, 
the  anomaly  (leading to  $ \langle W_L  \rangle \not= \langle W_C  \rangle$)
comes from a non-trivial  transformation of the  gauge vector  propagator under the  inversions
(and, hence, under  the special conformal transformation  in \rf{d}). 
Since in the above  example  the vector  propagators  did not contribute,   we should get  the same answer for 
both the  line and the circle.}
%At higher orders  in $\lambda$ gluons   may still 
%lead to   non-trivial  contributions so it is not a priori clear 
%if the  anomaly will be absent in the normalized  correlators like \rf{D.7}.
Similar agreement should be present    also  at higher orders in $\l$ and for more general correlators.

%This expression is  obviously  different
% from   the corresponding  one  \rf{3.31}
%for the  circular Wilson loop (interestingly, \rf{D16} can be obtained from 
%\rf{3.31}  if  we discard  in the 
%latter  the logarithmic and dilogarithmic terms
% in ${\cal F}$ in \rf{3.24}).
%This 
% suggests that  just as for the 
%expectation   values of $W_C$ and $W_L$  there is no  simple  relation between their 
%correlators with local  operators. 

\iffalse

{\bf  ???????????????????????
 
 1. what happens   for 1-point functions at weak coupling --  compare L and C ?
 
 2.  how  conf  transformation \rf{d}   acts on the $\u$, $\v$ variables  -- 
 they are actually same  in the two cases   so should be invariant 
 
 3.  discuss  strong coupling  -- take    line, $z= \sigma, \ x= \tau$   and compute 
 expectation value of say dilaton operator and compare with circle. 
 there will be difference. 
 
 4. what about integrated dilaton operator -- why 1-point function =0 in C case and 0 in L case then ?
 } 
 
 \fi

\iffalse 

\section{Conclusions}

normalization of N; 
anomaly circle line; 
no anomaly with operators.
comparison with GP

$\l^2$ order 

possibility to compute exactly if operators are in orthogonal plane 
(cf Semenoff-Zarembo)

contribution of internal vertices? 

\fi

%%%%%%%%%%%%%%%%%%%%%%%%%%%%%%%%%%%%%%%%%%%%%%%%%%%%%%%%%%%%%%%%%%%%%%%%%%%%%%%%%%%%%%%%%%%%%%%%%%%%%%%%%%%%%%%%%%%%%

\section*{Acknowledgments}

%%%%%%%%%%%%%%%%%%%%%%%%%%%%%%%%%%%%%%%%%%%%%%%%%%%%%%%%%%%%%%%%%%%%%%%%%%%%%%%%%%%%%%%%%%%%%%%%%%%%%%%%%%%%%%%%%%%%%%
We are grateful to N. Drukker,  S. Giombi, R. Roiban  and K. Zarembo  for very useful suggestions and 
 discussions.
This  work was  supported by   STFC grant ST/J000353/1.   
A.A.T. was  also  supported by the ERC Advanced  grant No.290456.

%%%%%%%%%%%%%%%%%%%%%%%%%%%%%%%%%%%%%%%%%%%%%%%%%%%%%%%%%%%%%%%%%%%%%%%%%%%%%%%%%%%%%%%%%%%%%%%%%%%%%%%%%%%%%%%%%%%%%
%\newpage
\appendix
\section{Infinitesimal conformal transformations\\ preserving 
the circle}
%%%%%%%%%%%%%%%%%%%%%%%%%%%%%%%%%%%%%%%%%%%%%%%%%%%%%%%%%%%%%%%%%%%%%%%%%%%%%%%%%%%%%%%%%%%%%%%%%%%%%%%%%%%%

Here we shall review the count of conformal symmetries 
preserved by the circle  before and after adding local operators (see 
  \ci{Bianchi:2002gz,Alday:2011pf}). 
 A general infinitesimal conformal 
transformation acts as follows
\be 
\d x_{\mu}= \a_{\mu} +\omega_{\mu \nu} x_{\nu} +\sigma  x_{\mu} +x^2 \b_{\mu} 
-2 (\b \cdot x) x_{\mu}\,,
\label{A1}
\ee
where the parameters $\a^{\mu}$, $\omega_{\mu \nu}$, $\sigma $ and $\beta^{\mu}$ correspond to 
translations, Lorentz transformations, dilatations and special conformal 
transformations respectively. 
Let us split $x^{\mu}=(x_1, x_2, x_3, x_4)$ into the components in the plane of the circle 
\eqref{1.12}
and in the orthogonal plane, $x_l= (x_1, x_2)$ and  $x_t= (x_3, x_4)$.
Below  we will fix the radius to be  $R=1$.
Taking into account that $x_l^2=1$, 
$x_t=0$ we get ($l,m=1,2$) 
\bea
\d x_l= \a_l +\omega_{l m} x_m +\sigma x_l + \b_l -(\b_m x_m)x_l\,, 
\ \ \ \ \ \ \
\d x_t= \a_t +\omega_{t m}x_m+\b_t\,.
\label{A2}
\eea
Now using $\d x_t=0$, $x_l \d x_l=0$ we obtain 
\be
\omega_{t m}=0\,, \quad  \sigma=0\,, \quad \a_t=-\b_t\,, \quad \a_l=\b_l\,.
\label{A3}
\ee
This means that the surviving  6 transformations are generated by
\be
\omega_{12}\,, \quad \omega_{34}\,, \quad \a_l\,, \quad \a_t\,, \quad 
\b_l=\a_l\,, \quad \b_t=-\a_t\,.
\label{A4}
\ee
An addition of an operator at a generic point of 4-space will break 4 out
of 6 conformal transformations~\eqref{A4}.   Introduction  of the second operator 
will break 
all the conformal transformations. 

%%%%%%%%%%%%%%%%%%%%%%%%%%%%%%%%%%%%%%%%%%%%%%%%%%%%%%%%%%%%%%%%%%%%%%%%%%%%%%%%%%%%%%%%%%%%%%%%%%%%%%%%%%%%%%%%%%%%%%%%%%%%%%

\section{Geodesic distances in $S^2$ and  $AdS_2$}

%%%%%%%%%%%%%%%%%%%%%%%%%%%%%%%%%%%%%%%%%%%%%%%%%%%%%%%%%%%%%%%%%%%%%%%%%%%%%%%%%%%%%%%%%%%%%%%%%%%%%%%%%%%%%%%%%%%%%%%%%%%%%%%%
Let us  present an analytic 
 derivation of the well-known expressions for the geodesic distances 
in $S^2$ and $AdS_2$ used in \rf{2.5}.

The geodesic $(\theta=\theta(t), \varphi=\varphi(t))$ on $S^2$ 
connecting the points 
$(\theta_1, \varphi_1)$  and $(\theta_2, \varphi_2)$ 
can be obtained by minimizing the functional 
\be
\ge =\int dt\  \big[(\partial_t \theta)^2 +\sin^2 \theta (\partial_t \varphi)^2\big]\,,
\label{B2}
\ee
 and evaluating it on the solution. 
 %Alternatively, we can eliminate $t$ and 
%consider the geodesic in the form $\theta=\theta(\varphi)$. Then the distance between the above points becomes
%
%\be
%\ell_1= \int_{\varphi_1}^{\varphi_2} \left[ \left(\frac{\partial \theta}{\partial \varphi}\right)^2+\sin^2 \theta\right]d \varphi\,.
%\label{B3}
%\ee
%
Integrating the  equations for  the geodesic gives  %followed from minimizing~\eqref{B2} are 
\bea
&&\partial_t(\sin^2 \theta \partial_t \varphi)=0\,, \qquad 
\partial^2_t \theta -\sin \theta \cos \theta (\partial_t \varphi)^2=0\,,
\label{B4} \\
&&
\cot \theta (t)=  \frac{C_2}{[C_1^2+ (C_1^2+C_2^2) \cot^2( \sqrt{C_1^2+C_2^2}(t-t_0)  )]^{1/2}}\,, 
\nonumber \\
&&
\cot (\varphi (t) -\varphi_0)= \frac{\sqrt{C_1^2+C_2^2}}{C_1} \cot( \sqrt{C_1^2+C_2^2} (t-t_0))\,,
\label{B5}
\eea
where $C_1, C_2, t_0, \varphi_0$ are 
 integration constants. Eliminating  $t$  we can  write the geodesic 
 passing  through the points $(\theta_1, \varphi_1)$ and $(\theta_2, \varphi_2)$
 in the form 
\bea
&&\cot \theta (\varphi)= A \sin \varphi +B \cos \varphi \,, \\
&&
A= \frac{\tan \theta_2 \cos \varphi_2- \tan \theta_1 \cos \varphi_1}{\tan \theta_2 \tan \theta_1 
\sin (\varphi_1-\varphi_2)}\,, \qquad 
B= \frac{\tan \theta_1 \sin \varphi_1- \tan \theta_2 \sin \varphi_2}{\tan \theta_2 \tan \theta_1 
\sin (\varphi_1-\varphi_2)}\,.
\label{B7}
\eea
Then the geodesic length   may  be written as 
 %integral~\eqref{B3} becomes
%
\bea
\ge= \int_{\varphi_1}^{\varphi_2}\frac{d \varphi\sqrt{1+A^2+B^2}}{1+ 
(A \sin \varphi +B \cos \varphi)^2}=
\arctan \frac{A B + (1+A^2) \tan \varphi_2}{\sqrt{1+A^2+B^2}}-  (\varphi_2 \to \varphi_1)
%\arctan \frac{A B + (1+A^2) \tan \varphi_1}{\sqrt{1+A^2+B^2}}
\,.
\label{B8}
\eea
As a result, one finds
\bea
\cos \ge
= \cos \theta_1 \cos \theta_2+ \sin \theta_1 \sin \theta_2 \cos(\varphi_2-\varphi_1)
\,.
\label{B12}
\eea
A similar analysis in $AdS_2$ gives for the corresponding geodesic distance $\ss$
\be
 \cosh \ss = \cosh \rho_1 \cosh \rho_2 -  \sinh \rho_1 \sinh \rho_2 \cos(\psi_2-\psi_1)\,.
\label{B14}
\ee
The two expressions are of course related  by the  analytic continuation
$\theta_k  \to i \rho_k$, $\varphi_k \to \psi_k$.

%%%%%%%%%%%%%%%%%%%%%%%%%%%%%%%%%%%%%%%%%%%%%%%%%%%%%%%%%%%%%%%%%%%%%%%%%%%%%%%%%%%%%%%%%%%%%%%%%%%%%%%%%%%%%%%%%%%%%%%%%%%%%%

\section{Correlator of circular Wilson loop with one light BPS  operator at 
strong coupling}

%%%%%%%%%%%%%%%%%%%%%%%%%%%%%%%%%%%%%%%%%%%%%%%%%%%%%%%%%%%%%%%%%%%%%%%%%%%%%%%%%%%%%%%%%%%%%%%%%%%%%%%%%%%%%%%%%%%%%%%%%%%%%%

Here we will review   the derivation of 
 correlation function of circular Wilson 
loop with one local 
operator which  will be chosen to be the dilaton or the  chiral primary
(with dimension $j$   fixed, i.e. not scaling with $\l$). 
 
The  correlator in question appeared in \rf{1.18},\rf{4.3}, 
i.e. in the leading large $\l$ approximation it is given by 
%According to our discussion in Section 4, we have 
%
\be 
{\cal C}(W_C, a)= \frac{\langle W_C {\ } {\cal O}(a)\rangle}{\langle W_C \rangle}=
\int d\tau d\s \  V(z(\tau, \s), x^{\mu}(\tau, \s)-a^{\mu}))\,,
\label{C1}
\ee
where $(z(\tau, \s), x^{\mu}(\tau, \s))$ 
represents the circular  loop solution~\eqref{4.0}. 
Since this correlator is fixed by conformal invariance up to a constant
 \rf{1.18},  we can
 choose 
the position  of the operator  to be at $a^{\mu}=(0, 0, \L, 0)$\,.

%The vertex operator of the dilaton is given in~\eqref{4.8.1}--\eqref{4.8.3}. 
Evaluating  the dilaton vertex operator
\rf{4.8.1} 
 on the solution~\eqref{4.10} gives (see \rf{1.14}) 
\be 
{\cal C}(W_C, a)= \frac{4 \pi \hat{c}_{dil}}{(L^2+1)^4} \int_0^{\infty}d \tau \
\frac{\sinh^2 \tau}{\cosh^4 \tau}=\frac{4 \pi  \hat{c}_{dil}}{3 (\L^2+1)^4}=
\frac{\pi \hat{c}_{dil}}{12} \frac{1}{[\tdr(a)]^4} \,.
\label{C2}
\ee
Using  the value of the  normalization coefficient~\eqref{4.8.3}
 we obtain %the coefficient 
${\rm C}_{dil}(\lambda)$  in~\eqref{4.6}.

The bosonic part of the chiral primary vertex operator 
of dimension $\D=j$ 
is given by
\ci{corr,zar11} 
\bea
\V(a)=\hat{c}_j \int d\tau d\s  \  \Big[\frac{z}{z^2+ (x^{\mu}-a^{\mu})^2}\Big]^{j} 
{\ }e^{i j \phi} {\ }U\,, 
\ \ \ \ \ \ \ \ \ \ \ 
\hat{c}_j=\frac{\sqrt{\lambda}}{8 \pi N} \sqrt{j} (j+1)\,,
\label{C3}
\eea
where  $\phi$ is the relevant angle in $S^1 \subset S^5$ 
and the two-derivative part $U$ is given by~\cite{btt} 
\bea
&&
U= U_1 +U_2+U_2\,, \ \ \ \ \ \ \ \ \ 
U_1=\frac{1}{z^2}[(\partial_ax^{\mu})^2 -(\partial_az)^2]-{\cal L}_{S^5}\,, 
%\label{C4}\\
\nonumber \\
&&
U_2=\frac{8}{[z^2 +(x^{\mu}-a^{\mu})^2]^2}
\Big[(x^{\mu}-a^{\mu})^2 (\partial_az)^2 - [(x^{\mu} -a^{\mu}) \partial_a x^{\mu}]^2\Big]\,, 
\nonumber\\
&&
U_3 =\frac{8 [(x^{\mu}-a^{\mu})^2 -z^2]}{z [z^2 +(x^{\mu} -a^{\mu})^2]^2}[(x_{\nu}-a_{\nu})\partial_ax_{\nu}]
\partial_az\,,
\label{C5}
\eea
where ${\cal L}_{S^5}$ is the $S^5$ part of the bosonic Lagrangian. 
Evaluating $U$ on the semiclassical Wilson loop 
background~\eqref{4.0}  (note that here $\phi=0$) gives 
\bea
U_1 =\frac{2}{\cosh^2 \tau}\,, 
 \ \ \ \ \ \ \ \ \ 
U_2=-U_3=\frac{8}{\L^2+1} \frac{1}{\cosh^6 \tau}
(\L^2 \cosh^2 \tau +1 -\sinh^2 \tau)\,.
\label{C6}
\eea
Thus  $U_2$ and $U_3$ cancel each other and we end up with 
\be
{\cal C}(W_C, a)=\frac{4 \pi \hat{c}_j}{(\L^2+1)^j}\int_0^{\infty}d \t \
\frac{\tanh^j \tau}{\cosh^2 \tau}=
\frac{1}{[\ell(a)]^j}\frac{\pi \hat{c}_j}{2^{j-2} (j+1)}\,.
\label{C7}
\ee
Using  the normalization $\hat{c}_j$ in~\eqref{C3} we find that  the coefficient
${\rm C}_j (\lambda)$ in \rf{1.18} is given by~\eqref{4.6}.

%%%%%%%%%%%%%%%%%%%%%%%%%%%%%%%%%%%%%%%%%%%%%%%%%%%%%%%%%%%%%%%%%%%%%%%%%%%%%%%%%%%%%%%%%%%%%%%%%%%%%%%%%%%%%%%%%%%%%%%%%%%%%%%%%%%%%%%%%%

\newpage

%\begin{thebibliography}{99}

\ed